\documentclass[%
reprint,
amsmath,amssymb,
aps,pra
]{revtex4-2}

\usepackage{amsthm}
\usepackage{graphicx}

\theoremstyle{plain}
\newtheorem{theorem}{Theorem}
\newtheorem{lemma}{Lemma}
\newtheorem{proposition}{Proposition}

\theoremstyle{definition}
\newtheorem{definition}{Definition}

\begin{document}

\title{Quantum Temporal Marginal Problem}
\author{Dingqian Wang}
\email{wangdingqian2001@gmail.com}
\affiliation{Thrust of Artificial Intelligence, Information Hub, The Hong Kong University of Science and Technology (Guangzhou), Guangzhou 511453, China}
\author{Jiayi Zhao}
\email{jzhao858@connect.hkust-gz.edu.cn}
\affiliation{Thrust of Artificial Intelligence, Information Hub, The Hong Kong University of Science and Technology (Guangzhou), Guangzhou 511453, China}
\author{Ge Bai}
\email{gebai@hkust-gz.edu.cn}
\affiliation{Thrust of Artificial Intelligence, Information Hub, The Hong Kong University of Science and Technology (Guangzhou), Guangzhou 511453, China}

\begin{abstract}
Experiments on a multi-time quantum process may reveal its responses only on selected, possibly overlapping, sets of times. We ask whether these partial descriptions admit a common underlying process, and formulate this process tensor marginal problem as a semidefinite feasibility test. Agreement on overlaps is necessary but insufficient: three locally consistent qubit marginals can have no common extension. Another family is incompatible as a time-ordered process even though a single quantum channel satisfies both its channel and temporal marginal constraints. This isolates temporal causality as the obstruction. We quantify incompatibility by the mixing needed to restore compatibility and derive finite tests using separate interventions at each time and final measurements. We further show that incompatibility is equivalent to an advantage over every compatible family in a multi-time process-discrimination task. Together, these results test whether process descriptions obtained in different experiments can arise from one causally ordered quantum process.
\end{abstract}

\maketitle

\section{Introduction}

Experiments probing quantum dynamics at several times often provide only partial descriptions of the underlying process. Different experiments may characterize its response to interventions on different, possibly overlapping sets of times. Before combining these descriptions into a dynamical model, one must determine whether they can arise from a common physical process. Process tensors describe these multi-time responses, including temporal correlations resulting from memory~\cite{Pollock:2018qgg,Milz:2020usa}. Complete reconstruction of the process is demanding: the experimentally available control operations may be restricted~\cite{milzReconstructingNonMarkovianQuantum2018,whiteDemonstrationNonMarkovianProcess2020,whiteWhatCanUnitary2025}, and the number of control sequences required for generic tomography grows exponentially with the number of intervention times~\cite{whiteDemonstrationNonMarkovianProcess2020,whiteNonMarkovianQuantumProcess2022}.

\begin{figure}[!tbp]
    \centering
    \includegraphics[width=\columnwidth]{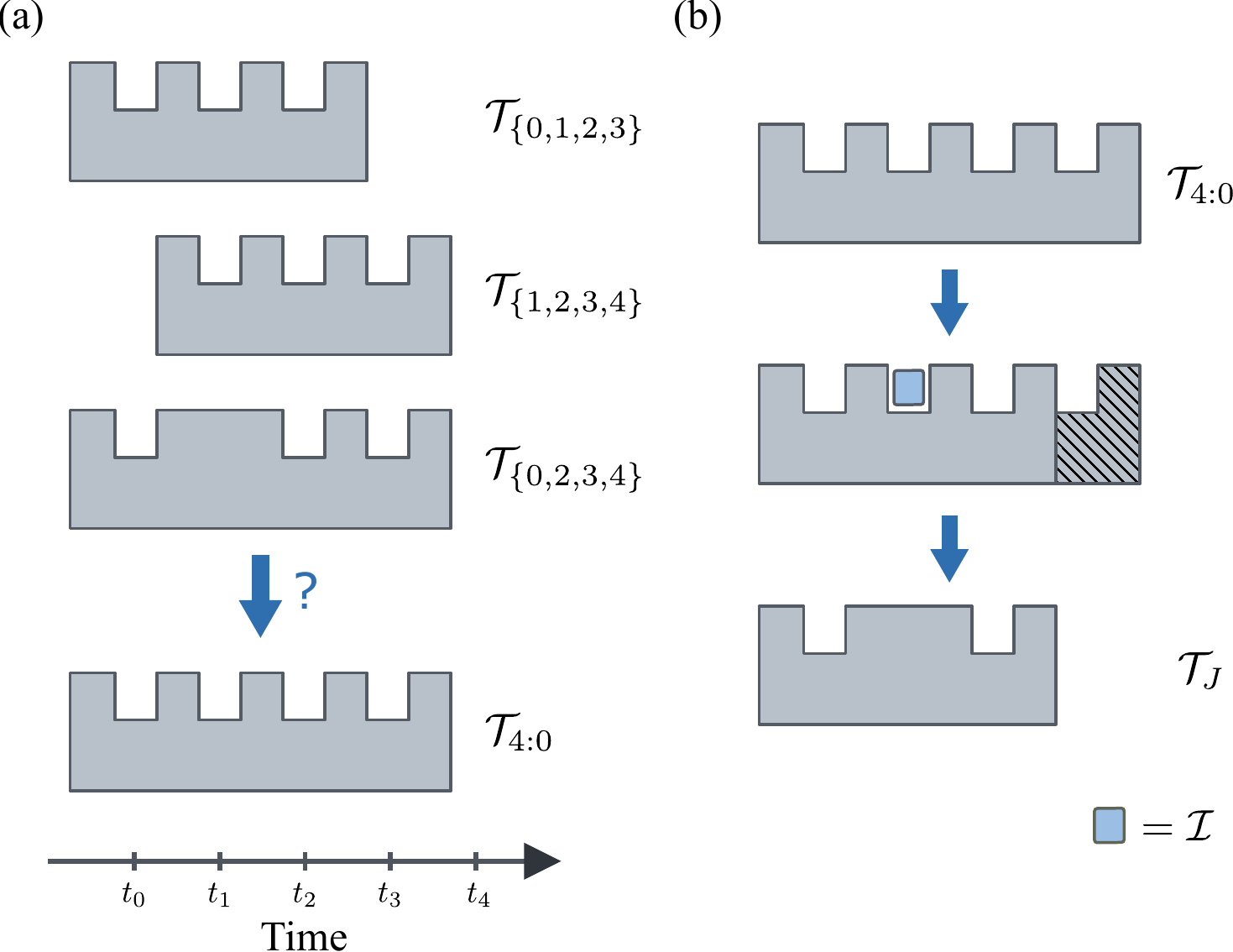}
    \caption{Process tensor marginal problem and temporal marginalization. Gray combs represent process tensors; slots between their teeth mark times at which controls may be applied. (a) Given several process descriptions on selected, possibly overlapping times, the task is to find one process on the full time sequence that reproduces all of them. (b) To obtain one temporal marginal, retain the process up to the final selected time. Selected slots before that time remain open for controls, while unselected slots are closed with identity channels (blue), representing no additional control. The hatched part lies after the final selected time.}
    \label{fig:problem-setting}
\end{figure}

We formulate the compatibility of these partial process descriptions as the \emph{process tensor marginal problem} (PTMP). Figure~\ref{fig:problem-setting}(a) depicts the task: several process descriptions cover different, possibly overlapping sets of times, and we seek one causally ordered process that reproduces all of them.

Figure~\ref{fig:problem-setting}(b) explains how a temporal marginal is obtained. The experimenter applies controls at selected times and reads out the system at the final selected time. Selected intervention slots before that readout remain open for those controls. At unselected times before the readout, no additional control is applied: the blue identity channels close those slots while the underlying evolution continues as usual~\cite{Pollock:2018qgg,berkExtractingQuantumDynamical2023}. The hatched part lies after the readout and is excluded from this marginal description.

This is a temporal counterpart of the quantum state and channel marginal problems. The state problem asks whether prescribed reductions arise by partial trace from one global state; agreement on shared subsystems is necessary but does not guarantee such a state~\cite{bravyiRequirementsCompatibilityLocal2004,klyachkoQuantumMarginalProblem2006}. An analogous extension question for quantum channels has also been studied~\cite{Hsieh:2021ewp}.

We first show that, with unrestricted finite-dimensional memory, valid marginals on disjoint selected time sets always admit a common extension, even when their temporal intervals interleave. Yet agreement on every overlap can still fail to yield a global process. In the qubit example of Sec.~\ref{sec:incompatible-example}, two nearest-neighbor marginals describe identity propagation, whereas an endpoint marginal describes bit-flip propagation. Their one-time marginals agree, yet no global process exists.

Temporal incompatibility can also persist despite the existence of a common quantum channel. In Sec.~\ref{sec:irreducibility-channelization}, we group the temporal interfaces of two valid marginals into channels and construct a global channel that reproduces them. The same channel also reproduces the marginals when omitted time slots are linked as in Fig.~\ref{fig:problem-setting}(b), yet no causally ordered process can do so. This isolates the constraint of temporal causality, even for classical stochastic behavior encoded in qubits. 

To test general families, we give an exact semidefinite formulation of compatibility and quantify incompatibility through the largest weight of the prescribed family retained in a compatible mixture. Building on convex witness and discrimination methods~\cite{Hsieh:2021ewp,takagiGeneralResourceTheories2019,hsiehResourceMarginalProblems2024}, we construct finite witnesses evaluated through separate completely positive interventions at individual times and final-output observables, without quantum memory in the controller. We also prove that a family is incompatible if and only if it outperforms every compatible family in some multi-time process-discrimination task, using the same process ensembles and final measurements throughout the comparison. These results provide operational tests of whether partial process descriptions admit a common dynamical model.

Section~\ref{sec:preliminaries} fixes the process-tensor notation. Section~\ref{sec:framework} defines temporal compatibility and derives its SDP formulation. Section~\ref{sec:irreducibility-channelization} establishes the separation from channel compatibility. Sections~\ref{sec:robustness} and~\ref{sec:operational-interpretation} develop robustness, witnesses, and the discrimination characterization. The appendices supply the proofs.

\section{Preliminaries}
\label{sec:preliminaries}

Let \(\mathcal L(\mathcal H)\) denote the space of linear operators on a Hilbert space \(\mathcal H\). We consider an open quantum process sampled at discrete time steps \(t_0<t_1<\cdots<t_k\). Following the process tensor framework of Pollock et al.~\cite{Pollock:2018qgg}, the process is described operationally by its response to interventions performed on the system.

This section reviews the Choi representation of process tensors, their action
on temporally correlated controls, and the recursive causal constraints used
throughout the compatibility and robustness analyses below.

At each time step \(j=0,\ldots,k-1\), the system immediately before the control operation is represented on an input Hilbert space \(\mathcal H_{i_j}\), and the system immediately after the control operation is represented on an output Hilbert space \(\mathcal H_{o_j}\). Thus a control operation at time \(j\) is a completely positive map
\begin{equation}
\mathcal A_j:\mathcal L(\mathcal H_{i_j})\to
\mathcal L(\mathcal H_{o_j}) .
\end{equation}
An individual outcome of an intervention is represented by a completely
positive (CP), trace-nonincreasing map, while a deterministic intervention is
represented by a completely positive and trace-preserving (CPTP) map. The
output of a process tensor is therefore subnormalized for a selected outcome
and normalized when every inserted control is deterministic.
Throughout, we assume that, at each intervention time \(j\),
\(\mathcal H_{i_j}\) and \(\mathcal H_{o_j}\) are fixed labeled
copies of the same system Hilbert space \(\mathcal H_j\), with
matched bases. Hence the identity channel
\(\mathcal I_j:\mathcal L(\mathcal H_{i_j})
\to\mathcal L(\mathcal H_{o_j})\) is well defined.
We write \(d_j:=\dim\mathcal H_j\); it may depend on \(j\).
We denote a sequence of controls by
\begin{equation}
\mathbf A_{k-1:0}:=
\{\mathcal A_{k-1},\ldots,\mathcal A_1,\mathcal A_0\}.
\end{equation}

A \(k\)-step process tensor is the map from such a sequence of control operations to the corresponding, possibly subnormalized, system state at time \(k\),
\begin{equation}
\rho_k:=\mathcal T_{k:0}[\mathbf A_{k-1:0}] .
\end{equation}
Thus \(\mathcal T_{k:0}\) characterizes the multi-time input-output behavior of the process on the time steps \(0,\ldots,k\). A deterministic process tensor is linear and completely positive and satisfies the causality, or containment, condition: the statistics available up to any time cannot depend on controls applied later~\cite{Pollock:2018qgg}. In the Choi representation, this condition becomes the recursive normalization in Eq.~\eqref{eq:pt-causal-trace-constraints}. Throughout, every process tensor is assumed deterministic, and a \emph{valid} process tensor means a deterministic one.

The final system at time \(k\) is represented on \(\mathcal H_{i_k}\). No control operation is applied at time \(k\), so \(\mathcal H_{i_k}\) is the Hilbert space of the final output state. We fix the ordering of the Hilbert-space factors as
\begin{equation}
\mathcal H_{k:0}:=
\mathcal H_{i_k}\otimes \mathcal H_{o_{k-1}}\otimes \mathcal H_{i_{k-1}}
\otimes\cdots\otimes \mathcal H_{o_0}\otimes \mathcal H_{i_0}.
\end{equation}

We use the Choi--Jamio{\l}kowski representation~\cite{jamiolkowskiLinearTransformationsWhich1972,choiCompletelyPositiveLinear1975} and adopt the unnormalized convention. For a control operation
\(\mathcal A_j\), define
\begin{equation}
J(\mathcal A_j)
:=
(\mathcal A_j\otimes \mathcal I_{i_j})
\bigl(|\Phi^+\rangle\langle\Phi^+|\bigr),
\qquad
|\Phi^+\rangle=\sum_r |r\rangle|r\rangle ,
\end{equation}
where \(|\Phi^+\rangle\) is the unnormalized maximally entangled state on
\(\mathcal H_{o_j}\otimes\mathcal H_{i_j}\), and \(\mathcal I_{i_j}\) denotes the identity map on \(\mathcal L(\mathcal H_{i_j})\). The Choi representation of the process tensor \(\mathcal T_{k:0}\) is denoted by
\begin{equation}
J(\mathcal T_{k:0}) := \Upsilon_{k:0}.
\end{equation}
More generally, a controller may retain an ancillary memory between
intervention times~\cite{gutoskiGeneralTheoryQuantum2007}. We represent the resulting temporally correlated sequence
as a CP multi-time control operation \(\mathbf A_{k-1:0}\), and denote its
Choi operator by \(J(\mathbf A_{k-1:0})\), ordered on
\begin{equation}
\mathcal H_{o_{k-1}}\otimes\mathcal H_{i_{k-1}}\otimes\cdots
\otimes\mathcal H_{o_0}\otimes\mathcal H_{i_0}.
\end{equation}
When the control operations at different time steps are independent,
the Choi operator factorizes as
\begin{equation}
J(\mathbf A_{k-1:0})
=
J(\mathcal A_{k-1})\otimes\cdots\otimes J(\mathcal A_0).
\end{equation}
In general, no such factorization is assumed. The action of the
process tensor is recovered from its Choi operator by
\begin{equation}
\mathcal T_{k:0}[\mathbf A_{k-1:0}]
=
\operatorname{Tr}_{o_{k-1}i_{k-1}\cdots o_0i_0}
\!\left[
\Upsilon_{k:0}
\bigl(
\mathbb I_{i_k}\otimes J(\mathbf A_{k-1:0})^T
\bigr)
\right],
\end{equation}
where \(\mathbb I_{i_k}\) is the identity operator on
\(\mathcal H_{i_k}\). The transpose is taken in the computational basis used
to define the Choi representation.

\section{Framework: process tensor marginal problem}
\label{sec:framework}

We now define the marginal process tensors used in this work. Following the standard construction of intermediate processes~\cite{Pollock:2018qgg}, we first restrict the global process to the final selected time. We then close every unselected earlier intervention slot with an identity channel and leave the selected earlier slots open. This operational marginalization differs from taking a partial trace of the global Choi operator.

In this section, we formalize this marginalization and the associated
compatibility problem. We prove that marginals on pairwise-disjoint selected
time sets are always compatible, even when their temporal intervals
interleave, and derive an SDP feasibility criterion for arbitrary finite
families.

\begin{definition}[Marginal process tensor]
Let \(\mathcal{T}_{k:0}\) be a process tensor on the time steps \(K=\{0,1,\ldots,k\}\), and let \(J=\{j_0<j_1<\cdots<j_n\}\subseteq \{0,1,\ldots,k\}\). Let \(j_n\) denote the final selected time, and define
\(J^-:=J\setminus\{j_n\}\). For \(n\geq1\), the marginal process tensor of \(\mathcal{T}_{k:0}\) on \(J\), denoted by \(\mathcal{T}_J\), is the map from operations at the times in \(J^-\) to the system state at time \(j_n\) given by
\begin{equation}
\mathcal{T}_J
\big[
\mathcal A_{j_{n-1}},\ldots,\mathcal A_{j_0}
\big]
:=
\mathcal{T}_{j_n:0}
\big[
\widetilde{\mathcal A}_{j_n-1},
\ldots,
\widetilde{\mathcal A}_0
\big].
\end{equation}
Here \(\mathcal{T}_{j_n:0}\) denotes the process tensor obtained by restricting \(\mathcal{T}_{k:0}\) to the interval \(\{0,\ldots,j_n\}\),
and
\begin{equation}
\widetilde{\mathcal A}_m
=
\begin{cases}
\mathcal A_m, & m\in J^-,\\
\mathcal I_m, & m\in \{0,\ldots,j_n-1\}\setminus J^-,
\end{cases}.
\end{equation}
Here \(m=0,\ldots,j_n-1\).
For a one-time set \(J=\{j_0\}\), there are no open intervention slots,
and the marginal is the state
\begin{equation}
\mathcal T_{\{j_0\}}
:=
\mathcal T_{j_0:0}
[\mathcal I_{j_0-1},\ldots,\mathcal I_0],
\end{equation}
with the bracket absent when \(j_0=0\).
This defines a marginalization map \(\mathcal M_J\) by
\begin{equation}
\mathcal{M}_J(\mathcal{T}_{k:0}) := \mathcal{T}_J .
\end{equation}
\end{definition}

Whenever \(\gamma\subseteq\alpha\), the notation
\(\mathcal M_\gamma(\mathcal T_\alpha)\) means that the same
marginalization construction is applied relative to the ambient
selected time set \(\alpha\).

\begin{proposition}
Let \(\varnothing\neq\gamma\subseteq\alpha\subseteq K\).
For every deterministic process tensor \(\mathcal T_{k:0}\),
\[
\mathcal M_\gamma\!\left(
    \mathcal M_\alpha(\mathcal T_{k:0})
\right)
=
\mathcal M_\gamma(\mathcal T_{k:0}).
\]
Moreover, \(\mathcal M_\alpha(\mathcal T_{k:0})\) is a
deterministic process tensor on \(\alpha\).
\end{proposition}

\begin{proof}
In the two-step marginalization, the first map inserts identity
operations at the times omitted from \(\alpha\), while the second
inserts them at the times in \(\alpha\) omitted from \(\gamma\).
Thus every intervention time preceding \(\max\gamma\) but not
belonging to \(\gamma\) is closed exactly once with the identity
operation. Causal restriction to the final time \(\max\gamma\)
is consistent by the containment property, proving the
composition identity. Since the inserted identity operations are
CPTP, linking them with a deterministic comb preserves complete
positivity and deterministic-comb normalization.
\end{proof}

We can now define the process tensor marginal problem. 

\begin{definition}[Process tensor marginal problem]
Consider a global set of time steps \(K=\{0,1,\ldots,k\}\) and a family of process tensors \(\{\mathcal T_\alpha\}_{\alpha\in\Lambda}\), where \(\Lambda\) is a finite collection of nonempty selected time sets \(\alpha\subseteq K\). The process tensor marginal problem asks whether there exists a global process tensor \(\mathcal T_{k:0}\) satisfying
\begin{equation}
\mathcal M_\alpha(\mathcal T_{k:0})
=
\mathcal T_\alpha,
\qquad
\forall \alpha\in\Lambda .
\label{eq:ptmp}
\end{equation}
\end{definition}

Any global extension must reproduce the same process on time steps shared by
two prescribed marginals. By the composition property above, this gives the
necessary overlap condition
\begin{equation}
\mathcal M_{\gamma}(\mathcal T_{\alpha})
=
\mathcal M_{\gamma}(\mathcal T_{\beta})
\quad\text{for}\quad
\gamma:=\alpha\cap\beta\neq\varnothing,
\qquad
\alpha,\beta\in\Lambda .
\end{equation}
A family of process tensors \(\{\mathcal T_{\alpha}\}_{\alpha\in\Lambda}\) satisfying this condition is said to be locally compatible.

\subsection{Nonoverlapping marginals are always compatible}

For the state marginal problem, and similarly for the channel marginal problem, nonoverlapping marginals form a trivial compatibility case: one can extend them independently on disjoint subsystems. For process tensors, the analogous statement requires slightly more care. Here ``nonoverlapping'' means disjointness of the selected time-step sets, not disjointness of the temporal intervals spanned by them. Thus two selected sets may be nonoverlapping even when their time spans interleave, for example \(J_1=\{0,2\}\) and \(J_2=\{1,3\}\). We show below that even in this interleaved situation, nonoverlapping process tensor marginals are still automatically compatible.

\begin{proposition}[Compatibility for pairwise-disjoint selected time sets]
Suppose that the selected time sets are pairwise disjoint, that is,
\begin{equation}
J_\alpha \cap J_\beta = \varnothing, \qquad \alpha \neq \beta .
\end{equation}
Then every family of valid marginal process tensors \(\{\mathcal{T}_{J_\alpha}\}_{\alpha\in\Lambda}\) is globally compatible.
\end{proposition}

\begin{proof}
If the selected sets are also separated in the global time order, in the sense that all times in one selected set occur before all times in another, the construction is immediate: place the corresponding marginal process tensors on independent time blocks, and on every remaining time step let the process prepare a fixed state on the input leg and discard the returned output leg. We call such a single-time process a \emph{dummy process}.

It remains to treat the general case in which the selected sets are disjoint as sets but interleaved in time. In this case, a compatible global process can still be constructed by assigning an independent memory register to each marginal process tensor. For each \(\alpha\), introduce a private memory register whose input-output behavior at the selected times in \(J_\alpha\) is described by the marginal process tensor \(\mathcal{T}_{J_\alpha}\). These private memory registers are then embedded into the same global time order. Whenever the global time \(t\) belongs to \(J_\alpha\), the input leg \(i_t\) is prepared according to the private memory associated with \(\mathcal{T}_{J_\alpha}\). If \(t\) is not the final selected time of \(J_\alpha\), then the output leg \(o_t\) returned by the control operation is fed back only to that same private memory. Whenever \(t\) does not belong to any selected set, the global process acts as a dummy process at that time step.

Because the sets \(J_\alpha\) are pairwise disjoint, no two private memories are ever required to act on the same input or output leg. The resulting object is therefore a valid global process tensor: the input leg \(i_t\) at each time \(t\) is prepared using only the private memory registers updated by earlier returned output legs.

By construction, marginalizing the global process to \(J_\alpha\) closes or removes all slots associated with the dummy process and with the private memories for \(J_\beta\), \(\beta\neq\alpha\), leaving only the private memory that implements \(\mathcal{T}_{J_\alpha}\). Therefore,
\begin{equation}
\mathcal M_{J_\alpha}(\mathcal{T}_{k:0}) = \mathcal{T}_{J_\alpha},
\qquad \forall \alpha\in\Lambda .
\end{equation}
This proves global compatibility.
\end{proof}

\begin{figure}[t]
    \centering
    \includegraphics[width=\linewidth, page=1]{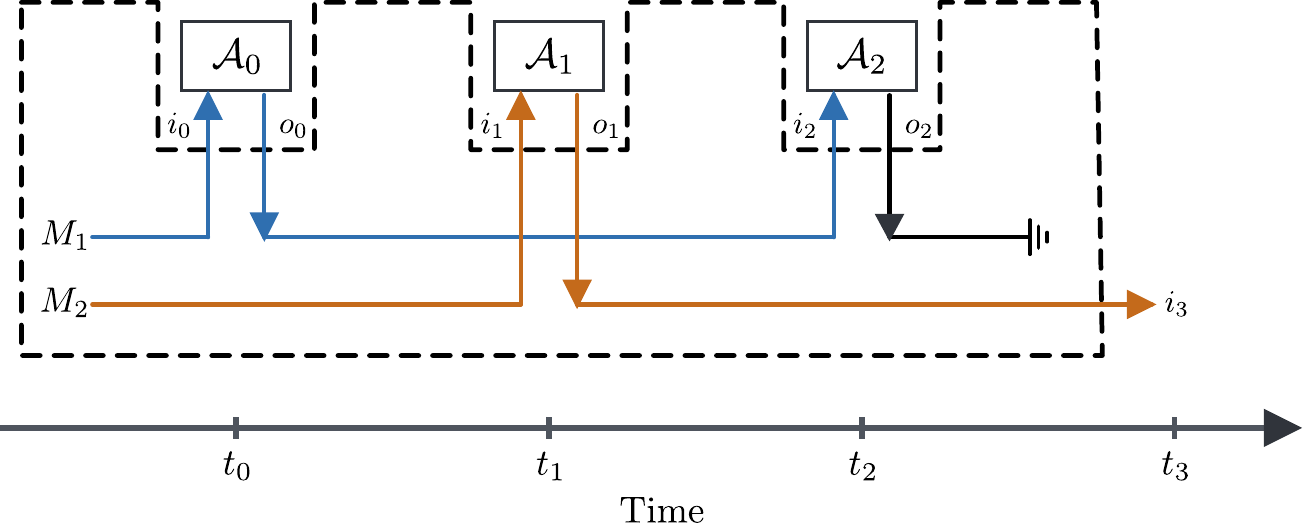} 
    \caption{Global process tensor construction for the interleaved but nonoverlapping marginals on $J_1=\{0,2\}$ and $J_2=\{1,3\}$.
    The dashed outer box represents the global comb, and the smaller dashed
    boxes mark its three intervention slots. The upper (blue) and lower
    (orange) lines represent the independent
    private memories $M_1$ and $M_2$, respectively, which are accessed
    at alternating time steps without exchanging information. At $t_2$,
    $M_1$ supplies $i_2$, whereas the returned output $o_2$ is discarded,
    leaving $M_2$ unaffected. The memory $M_2$ subsequently produces the
    final state $i_3$ at $t_3$. The resulting global comb
    $\mathcal T_{3:0}$ satisfies
    $\mathcal M_{J_k}(\mathcal T_{3:0})=\mathcal T_{J_k}$ for $k=1,2$.}
    \label{fig:nonoverlapping_marginals_example}
\end{figure}

Consequently, incompatibility in the PTMP requires overlaps among selected
time sets and arises from the interaction between their overlap constraints
and the global temporal order. Figure~\ref{fig:nonoverlapping_marginals_example}
illustrates the interleaved pair \(J_1=\{0,2\}\) and \(J_2=\{1,3\}\).

\subsection{PTMP as a semidefinite program}
\label{sec:iiib}

We formulate the PTMP in the Choi representation. By abuse of notation, we use \(\mathcal M_\alpha\) both for the marginalization map on process tensors and for the induced linear map on their Choi operators, i.e.,
\begin{equation}
\mathcal M_\alpha(\Upsilon_{k:0})
:=
J\!\left(\mathcal M_\alpha(\mathcal T_{k:0})\right),
\qquad
\Upsilon_{k:0}=J(\mathcal T_{k:0}) .
\end{equation}
For each prescribed marginal process tensor \(\mathcal T_\alpha\), we denote its Choi operator by \(\Upsilon_\alpha:=J(\mathcal T_\alpha)\).

Let \(\Lambda\) be the collection of prescribed selected time sets.
For a fixed \(\alpha\in\Lambda\), write its selected
time steps as
\(\alpha=\{j_1<j_2<\cdots<j_n\}\subseteq\{0,\ldots,k\}\).
Let
\begin{equation}
\beta
:=
\{0,\ldots,j_n-1\}\setminus\alpha
=
\{\ell_0<\ell_1<\cdots<\ell_m\}
\end{equation}
be the set of unselected intervention times preceding the final
selected time \(j_n\). If this set is empty, all products and partial traces
over its elements below are understood to be absent.

Let \(\Upsilon_{j_n:0}\) denote the Choi operator of the process tensor
obtained by restricting \(\mathcal T_{k:0}\) to the interval
\(\{0,\ldots,j_n\}\). In the unnormalized Choi convention used here,
it is given by
\begin{equation}
\Upsilon_{j_n:0}
=
\left(
\prod_{r=j_n}^{k-1}\frac{1}{d_{o_r}}
\right)
\operatorname{Tr}_{i_k,o_{k-1},\ldots,i_{j_n+1},o_{j_n}}
\left[
\Upsilon_{k:0}
\right].
\end{equation}
Here and below, an empty product equals one. For an intervention time \(m\),
write
\begin{equation}
\Phi_m^+
:=
|\Phi_m^+\rangle\langle\Phi_m^+|,
\qquad
|\Phi_m^+\rangle
:=
\sum_r |r\rangle_{o_m}|r\rangle_{i_m},
\end{equation}
so that \(\Phi_m^+=J(\mathcal I_m)\) in our unnormalized convention.
Adapting Eq.~(19) of Ref.~\cite{Pollock:2018qgg} to the unnormalized
Choi convention used here, the Choi representation of the
marginalization map is given by
\begin{equation}
\begin{aligned}
\mathcal M_{\alpha}(\Upsilon_{k:0})
&=
\operatorname{Tr}_{\{i_{\ell_r},o_{\ell_r}\}_{r=0}^{m}}
\\[-1mm]
&\quad
\Bigl[
\left(
\Phi^+_{\ell_m}\otimes\cdots\otimes
\Phi^+_{\ell_1}\otimes\Phi^+_{\ell_0}
\right)
\Upsilon_{j_n:0}
\Bigr].
\end{aligned}
\end{equation}

We now obtain a semidefinite formulation of the PTMP. The validity conditions for the unknown global process tensor are
positivity and the linear constraints for deterministic quantum combs:
\begin{equation}
\operatorname{Tr}_{i_j}(\Upsilon_{j:0})
=
\mathbb I_{o_{j-1}}\otimes \Upsilon_{j-1:0},
\; j=1,\ldots,k;
\quad
\operatorname{Tr}(\Upsilon_{0:0})=1 .
\label{eq:pt-causal-trace-constraints}
\end{equation}
Together, positivity and these linear constraints characterize the Choi operators of deterministic quantum combs~\cite{chiribellaQuantumCircuitArchitecture2008,
chiribellaTheoreticalFrameworkQuantum2009}. The base condition \(\operatorname{Tr}(\Upsilon_{0:0})=1\) corresponds to the initial-state normalization, equivalently to a comb normalization with a one-dimensional initial input wire. This leads to the following semidefinite feasibility problem:
\begin{equation}
\begin{aligned}
\text{find}\quad
& \{\Upsilon_{j:0}\}_{j=0}^k \\[1mm]
\text{subject to}\quad
& \Upsilon_{j:0} \succeq 0,
&& j=0,\ldots,k, \\[1mm]
& \operatorname{Tr}(\Upsilon_{0:0})=1, \\[1mm]
& \operatorname{Tr}_{i_j}(\Upsilon_{j:0})
=
\mathbb I_{o_{j-1}}\otimes \Upsilon_{j-1:0},
&& j=1,\ldots,k, \\[1mm]
& \mathcal M_\alpha(\Upsilon_{k:0})=\Upsilon_\alpha,
&& \forall \alpha\in\Lambda .
\end{aligned}
\label{eq:ptmp-sdp}
\end{equation}

\section{Robustness and incompatibility witnesses}
\label{sec:robustness}

Robustness measures quantify how much of an incompatible object can be
retained while mixing it with another valid object to restore compatibility.
They also lead to separating witnesses and operational characterizations in
input-output tasks~\cite{carmeliWitnessingIncompatibilityQuantum2019,carmeliQuantumIncompatibilityWitnesses2019,hallCompatibilitySubsystemStates2007,rossetResourceTheoryQuantum2018,uolaQuantificationQuantumDynamics2020,skrzypczykAllSetsIncompatible2019,moriOperationalCharacterizationIncompatibility2020,yuanUniversalOperationalBenchmarking2021,takagiGeneralResourceTheories2019,uolaQuantifyingQuantumResources2019}. Related constructions have been studied for quantum devices, state and channel marginal problems, testers of multi-step causal networks, and multi-time processes~\cite{haapasaloRobustnessIncompatibilityQuantum2015,haapasaloQuantumMarginalProblem2021,Hsieh:2021ewp,hsiehResourceMarginalProblems2024,sedlakIncompatibleMeasurementsQuantum2016,berkResourceTheoriesMultitime2021}. We now define the process-tensor analogue of the family-wise channel incompatibility robustness in Ref.~\cite{Hsieh:2021ewp}. We formulate this robustness as an SDP, establish strong duality, and derive separating witnesses for every incompatible family. We further show that these witnesses can be evaluated using finitely many CP multi-time controls and final-output observables, and provide an explicit family that is locally compatible but globally incompatible.

\begin{definition}[Process tensor incompatibility robustness]
Let \(\boldsymbol{\mathcal T}:=\{\mathcal T_\alpha\}_{\alpha\in\Lambda}\) be the input family of marginal process tensors, and let \(\mathfrak C\) denote the set of families of marginal process tensors that admit a common global process tensor. We define the incompatibility robustness by
\begin{equation}
R(\boldsymbol{\mathcal T})
:=
\max
\left\{
\lambda\in[0,1]
\;\middle|\;
\lambda \boldsymbol{\mathcal T}
+
(1-\lambda)\boldsymbol{\mathcal N}
\in
\mathfrak C
\right\},
\label{eq:ptmp-robustness}
\end{equation}
where the maximization is over families of valid marginal process tensors \(\boldsymbol{\mathcal N}:=\{\mathcal N_\alpha\}_{\alpha\in\Lambda}\). The linear combination of \(\boldsymbol{\mathcal T}\) and \(\boldsymbol{\mathcal N}\) is defined componentwise, that is, \(a\boldsymbol{\mathcal T}+b\boldsymbol{\mathcal N}:=\{a\mathcal T_\alpha+b\mathcal N_\alpha\}_{\alpha\in\Lambda}\).
\end{definition}

\begin{minipage}{\dimexpr\columnwidth-10pt\relax}
For fixed finite-dimensional leg spaces and a fixed collection \(\Lambda\),
the compatible set \(\mathfrak C\) is compact and convex in the Choi
representation. Indeed, the global deterministic-comb Choi operators form a
closed convex set of positive operators with fixed trace, and hence a compact
set. The joint marginalization map is linear and continuous, so its image
\(\mathfrak C\) is compact and convex. Consequently, every continuous linear
score attains its maximum over \(\mathfrak C\). The sets of valid marginal
Choi operators are likewise compact, making the feasible set of
Eq.~\eqref{eq:ptmp-robustness}, in the variables \(\lambda\) and
\(\boldsymbol{\mathcal N}\), compact. Thus the maximum defining
\(R(\boldsymbol{\mathcal T})\) is attained, and
\(R(\boldsymbol{\mathcal T})=1\) holds exactly when
\(\boldsymbol{\mathcal T}\in\mathfrak C\).
\end{minipage}

By construction, \(R(\boldsymbol{\mathcal T})\) is the largest common weight of the prescribed family that can be retained while mixing it with another valid family so that the resulting family admits a common global process-tensor extension. Consequently, \(R(\boldsymbol{\mathcal T})=1\) if and only if \(\boldsymbol{\mathcal T}\in\mathfrak C\), whereas a smaller value means that a larger fraction of the prescribed family must be replaced in order to restore compatibility. This definition is closely related to robustness measures used for quantum-device, marginal, and channel incompatibility~\cite{haapasaloRobustnessIncompatibilityQuantum2015,haapasaloQuantumMarginalProblem2021,Hsieh:2021ewp}.

For each \(\mathcal T_\alpha\), we denote its Choi operator by \(\Upsilon_\alpha\). Then \(R(\boldsymbol{\mathcal T})\) is the solution of the following semidefinite program (SDP):
\begin{equation}
\begin{aligned}
\max_{\lambda,\{\Upsilon_{j:0}\}_{j=0}^{k}}
\quad
& \lambda \\[1mm]
\mathrm{subject\ to}
\quad
& 0\le \lambda\le 1, \\[1mm]
& \Upsilon_{j:0}\succeq 0,
&& j=0,\ldots,k, \\[1mm]
& \operatorname{Tr}(\Upsilon_{0:0})=1, \\[1mm]
& \operatorname{Tr}_{i_j}(\Upsilon_{j:0})
=
\mathbb I_{o_{j-1}}\otimes \Upsilon_{j-1:0},
&& j=1,\ldots,k, \\[1mm]
& \mathcal M_{\alpha}(\Upsilon_{k:0})
\succeq
\lambda\,\Upsilon_\alpha,
&& \forall \alpha\in\Lambda .
\end{aligned}
\label{eq:ptmp-robustness-sdp}
\end{equation}
The SDP in Eq.~\eqref{eq:ptmp-robustness-sdp} is obtained from the compatibility SDP in Eq.~\eqref{eq:ptmp-sdp} by relaxing the exact marginal equalities to inequalities with a common scaling parameter \(\lambda\) and maximizing \(\lambda\). It therefore provides both a compatibility test and a quantitative characterization of incompatibility. Moreover, an optimal solution specifies a compatible global process tensor whose marginals realize the corresponding optimal mixture.

As in other convex compatibility and resource problems, the dual program yields separating witnesses for families lying outside the compatible set~\cite{hallCompatibilitySubsystemStates2007,carmeliWitnessingIncompatibilityQuantum2019,carmeliQuantumIncompatibilityWitnesses2019,uolaQuantifyingQuantumResources2019,Hsieh:2021ewp}. The dual of the robustness SDP is
\begin{equation}
\begin{aligned}
\min_{\substack{
h,z,\{H_j\}_{j=1}^k,\\
\{Z_\alpha\}_{\alpha\in\Lambda}
}}
\quad
& h+z \\[1mm]
\mathrm{subject\ to}
\quad
& h\mathbb I_{i_0}
-
\operatorname{Tr}_{o_0}(H_1)
\succeq 0, \\[1mm]
& \mathbb I_{i_j}\otimes H_j
-
\operatorname{Tr}_{o_j}(H_{j+1})
\\[-1mm]
& \qquad \succeq 0,
\quad j=1,\ldots,k-1, \\[1mm]
& \mathbb I_{i_k}\otimes H_k
-
\sum_{\alpha\in\Lambda}
\mathcal M_{\alpha}^{\dagger}(Z_\alpha)
\succeq 0, \\[1mm]
& z+
\sum_{\alpha\in\Lambda}
\operatorname{Tr}\!\left(Z_\alpha\Upsilon_\alpha\right)
\geqslant 1, \\[1mm]
& H_j^\dagger=H_j,
\quad j=1,\ldots,k, \\[1mm]
& h\in\mathbb R, \\[1mm]
& Z_\alpha\succeq 0,
\quad \forall \alpha\in\Lambda, \\[1mm]
& z\geqslant 0 .
\end{aligned}
\label{eq:ptmp-robustness-dual}
\end{equation}
The dual formulation and strong duality are proved in Appendix~\ref{app:ptmp-robustness-dual}. In particular, strong duality follows from Slater's condition: the primal problem is feasible and bounded, while the dual problem is strictly feasible. Hence, Eqs.~\eqref{eq:ptmp-robustness-sdp} and~\eqref{eq:ptmp-robustness-dual} have the same optimum.

Let \(\boldsymbol{\Upsilon}^{\mathcal T}:=\{\Upsilon_\alpha^{\mathcal T}\}_{\alpha\in\Lambda}\) be the family of Choi operators associated with \(\boldsymbol{\mathcal T}\). For a family of operators \(\mathbf A:=\{A_\alpha\}_{\alpha\in\Lambda}\) and any family of Choi operators \(\boldsymbol{\Upsilon}:=\{\Upsilon_\alpha\}_{\alpha\in\Lambda}\), define
\begin{equation}
\langle \mathbf A,\boldsymbol{\Upsilon}\rangle
:=
\sum_{\alpha\in\Lambda}
\operatorname{Tr}\!\left(
A_\alpha\Upsilon_\alpha
\right).
\label{eq:choi-family-pairing}
\end{equation}

\begin{proposition}[Choi incompatibility witness]
\label{prop:choi-incompatibility-witness}
The family \(\boldsymbol{\mathcal T}\) is incompatible if and only if there exists a family of positive operators \(\boldsymbol W:=\{W_\alpha\}_{\alpha\in\Lambda}\), with \(W_\alpha\succeq 0\) for every \(\alpha\in\Lambda\), such that
\begin{equation}
\langle
\boldsymbol W,
\boldsymbol{\Upsilon}^{\mathcal T}
\rangle
>
\max_{\boldsymbol{\Gamma}\in\mathfrak C}
\langle
\boldsymbol W,
\boldsymbol{\Upsilon}^{\Gamma}
\rangle .
\label{eq:choi-incompatibility-witness}
\end{equation}
\end{proposition}

The proof of Proposition~\ref{prop:choi-incompatibility-witness} is given in Appendix~\ref{app:choi-incompatibility-witness}. Proposition~\ref{prop:choi-incompatibility-witness} is the process-tensor counterpart of incompatibility witnesses previously developed for subsystem states, measurements, quantum channels, and channel marginal problems~\cite{hallCompatibilitySubsystemStates2007,carmeliQuantumIncompatibilityWitnesses2019,carmeliWitnessingIncompatibilityQuantum2019,Hsieh:2021ewp}. It characterizes process-tensor incompatibility using positive operators on the Choi spaces of the marginal process tensors. The following result is specific to the multi-time setting: it represents such witnesses using finite collections of multi-time control operations, which may themselves be temporally correlated.

\begin{theorem}[Process tensor incompatibility witness]
\label{thm:process-tensor-incompatibility-witness}
The family \(\boldsymbol{\mathcal T}\) is incompatible if and only if, for each \(\alpha\in\Lambda\), there is a finite collection of Hermitian operators \(\{F_\alpha^{(i)}\}_i\) and CP multi-time control operations \(\{\mathbf A_\alpha^{(i)}\}_i\), possibly temporally correlated, such that
\begin{equation}
\begin{aligned}
&
\sum_{\alpha\in\Lambda}
\sum_i
\operatorname{Tr}\!\left[
F_\alpha^{(i)}
\,
\mathcal T_\alpha
\!\left[
\mathbf A_\alpha^{(i)}
\right]
\right]
\\
&\quad
>
\max_{\boldsymbol{\Gamma}\in\mathfrak C}
\sum_{\alpha\in\Lambda}
\sum_i
\operatorname{Tr}\!\left[
F_\alpha^{(i)}
\,
\Gamma_\alpha
\!\left[
\mathbf A_\alpha^{(i)}
\right]
\right].
\end{aligned}
\label{eq:operational-incompatibility-witness}
\end{equation}
\end{theorem}

The proof of Theorem~\ref{thm:process-tensor-incompatibility-witness} is given in Appendix~\ref{app:process-tensor-incompatibility-witness}. The finiteness in the theorem makes the separation a finite certificate of incompatibility. Although Proposition~\ref{prop:choi-incompatibility-witness} initially represents the witness by positive operators on the Choi spaces, each such operator can be represented using finitely many completely positive multi-time control operations and finitely many Hermitian operators on the corresponding final-output spaces. Consequently, certifying incompatibility requires only a finite family of interventions. The operators \(F_\alpha^{(i)}\) need not themselves be measurement effects; their conversion into a physically valid discrimination task is established in Sec.~\ref{sec:operational-interpretation}.

\subsection{Example of incompatibility}
\label{sec:incompatible-example}

We give a simple example showing that local compatibility of process tensor
marginals does not imply global compatibility. Consider a qubit process on
the time set \(K=\{0,1,2\}\). Let \(\rho_0=\mathbb I/2\), and let
\(\mathcal X\) be the bit-flip channel, \(\mathcal X(\rho)=X\rho X\), where
\(X\) is the Pauli bit-flip operator.

We prescribe three one-step marginal process tensors on
\(J_1=\{0,1\}\), \(J_2=\{1,2\}\), and \(J_3=\{0,2\}\) by
\begin{equation}
\mathcal T_{J_1}[\mathcal A_0]
=
\mathcal A_0(\rho_0),
\end{equation}
\begin{equation}
\mathcal T_{J_2}[\mathcal A_1]
=
\mathcal A_1(\rho_0),
\end{equation}
and
\begin{equation}
\mathcal T_{J_3}[\mathcal A_0]
=
\mathcal X\!\left(\mathcal A_0(\rho_0)\right).
\label{eq:main-example-TJ3}
\end{equation}
Thus the nearest-neighbor marginals \(J_1\) and \(J_2\) describe identity
propagation, while the non-neighbor marginal \(J_3\) describes bit-flip
propagation from time \(0\) to time \(2\).

\begin{figure}[t]
    \centering
    \includegraphics[width=\columnwidth]{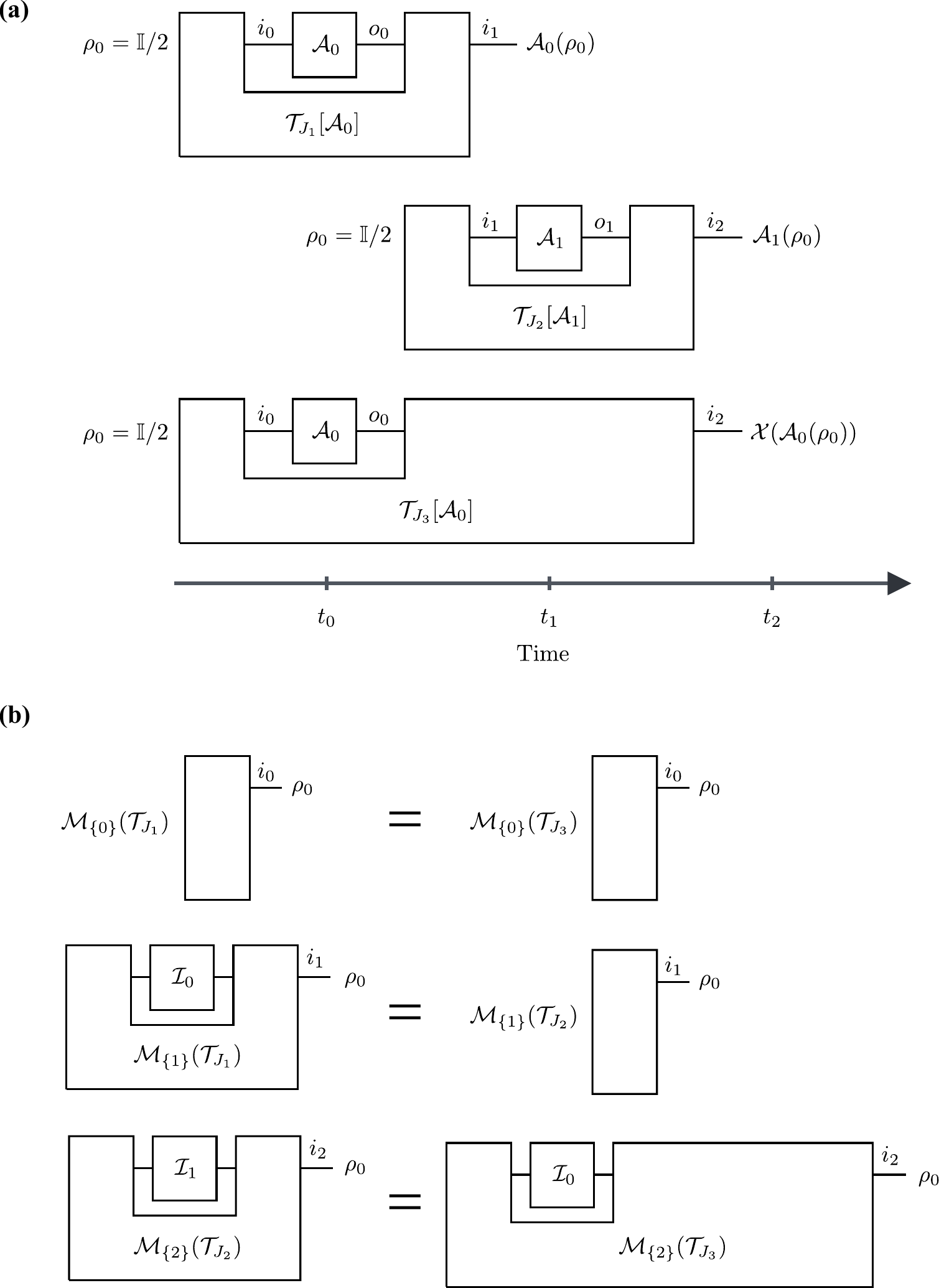}
    \caption{Locally compatible but globally incompatible process tensor marginals. (a) Three one-step qubit process tensors on $J_1=\{0,1\}$, $J_2=\{1,2\}$, and $J_3=\{0,2\}$, with $\rho_0=\mathbb I/2$. The nearest-neighbor marginals $\mathcal T_{J_1}$ and $\mathcal T_{J_2}$ describe identity propagation, whereas $\mathcal T_{J_3}$ applies the bit-flip channel $\mathcal X(\rho)=X\rho X$ between $t_0$ and $t_2$. (b) The three process tensors agree on every pairwise overlap: closing the appropriate intervention slots with identity channels yields the same maximally mixed state $\rho_0$. Nevertheless, no global process tensor $\mathcal T_{2:0}$ has all three as temporal marginals.}
    \label{fig:incompatibility-example}
\end{figure}

These three marginals are locally compatible. Indeed, the pairwise overlaps
are \(J_1\cap J_2=\{1\}\), \(J_1\cap J_3=\{0\}\), and
\(J_2\cap J_3=\{2\}\). On each overlap, the corresponding one-time marginal
is the maximally mixed state \(\rho_0\). In particular,
\(\mathcal X(\rho_0)=X\rho_0X=\rho_0\). Hence all pairwise overlap
constraints are satisfied.
The three marginals and their pairwise overlap relations are illustrated in
Fig.~\ref{fig:incompatibility-example}.

However, the family
\(\boldsymbol{\mathcal T}
=
\{\mathcal T_{J_1},\mathcal T_{J_2},\mathcal T_{J_3}\}\)
is not globally compatible. Suppose, for contradiction, that there exists a
global two-step process tensor \(\mathcal T_{2:0}\) such that
\begin{equation}
\mathcal M_{J_a}(\mathcal T_{2:0})
=
\mathcal T_{J_a},
\qquad a=1,2,3.
\end{equation}

As shown in Appendix~\ref{app:locally-compatible-incompatible}, the first two marginal constraints uniquely determine the Choi
operator of the assumed global extension. In particular, writing
\(\Upsilon_{2:0}:=J(\mathcal T_{2:0})\), one obtains
\begin{equation}
\Upsilon_{2:0}
=
\Phi^+_{i_2o_1}
\otimes
\Phi^+_{i_1o_0}
\otimes
(\rho_0)_{i_0}.
\end{equation}
Thus, closing the time-\(1\) slot with the identity control gives
\begin{equation}
\mathcal M_{J_3}(\Upsilon_{2:0})
=
\Phi^+_{i_2o_0}\otimes(\rho_0)_{i_0}.
\end{equation}
Hence the induced channel from time \(0\) to time \(2\) must be the
identity channel, so that
\begin{equation}
\mathcal M_{J_3}(\mathcal T_{2:0})[\mathcal A_0]
=
\mathcal A_0(\rho_0)
\end{equation}
for every admissible operation \(\mathcal A_0\).

This contradicts the prescribed marginal on \(J_3\), which requires
Eq.~\eqref{eq:main-example-TJ3}. For example, if \(\mathcal A_0\) is
the replacement operation preparing \(\lvert0\rangle\langle0\rvert\),
then any global extension would give
\begin{equation}
\mathcal M_{J_3}(\mathcal T_{2:0})[\mathcal A_0]
=
\lvert0\rangle\langle0\rvert,
\end{equation}
whereas the prescribed marginal gives
\begin{equation}
\mathcal T_{J_3}[\mathcal A_0]
=
X\lvert0\rangle\langle0\rvert X
=
\lvert1\rangle\langle1\rvert.
\end{equation}
This contradiction shows that the locally compatible family
\(\boldsymbol{\mathcal T}\) admits no common global process tensor
extension.

For this family, solving the robustness SDP in
Eq.~\eqref{eq:ptmp-robustness-sdp} numerically gives
\(R(\boldsymbol{\mathcal T})\approx 0.75\) to numerical precision.

\section{Irreducibility to the channel marginal problem under channelization}
\label{sec:irreducibility-channelization}

We first define the channelization used in this section. For a selected time
set \(J=\{j_0<\cdots<j_n\}\), define
\begin{equation}
\mathcal H_{I_J}
:=
\mathcal H_{i_{j_n}}\otimes\cdots\otimes\mathcal H_{i_{j_0}}.
\end{equation}
For \(n\geq1\), let
\begin{equation}
\mathcal H_{O_J}
:=
\mathcal H_{o_{j_{n-1}}}\otimes\cdots\otimes\mathcal H_{o_{j_0}},
\end{equation}
and for a one-time set \(J=\{j_0\}\), let
\(\mathcal H_{O_J}:=\mathbb C\).
The recursive normalization of a deterministic process tensor implies
\begin{equation}
\operatorname{Tr}_{I_J}\Upsilon_J
=
\mathbb I_{O_J}.
\end{equation}
Thus \(\Upsilon_J\) is also the Choi operator of a CPTP map from
\(\mathcal H_{O_J}\) to \(\mathcal H_{I_J}\). We call this channel
\(\operatorname{Ch}(\mathcal T_J)\) the \emph{channelization} of
\(\mathcal T_J\). Channelization retains overall complete positivity and
trace preservation but discards the intermediate recursive causal
constraints. Compatibility of the resulting channels is understood in the
channel-marginal sense, including the no-signaling conditions required for
their reduced channels to be well defined~\cite{Hsieh:2021ewp}.

Channelization identifies a deterministic process-tensor Choi operator with the Choi operator of a CPTP map on grouped input and output spaces. It also changes the marginal relation: temporal marginalization closes omitted intervention slots with identity channels, whereas an ordinary channel marginal is defined by tracing out outputs and imposing independence from the removed inputs. These two operations do not generally agree. We show that compatibility of the channelized family is insufficient for a common temporal extension.

It is therefore natural to ask whether PTMP compatibility can be tested by
channelizing each prescribed marginal and solving a channel
marginal problem. In this section, we prove that this reduction is not valid
in general: channelized marginals can admit a common global channel even when
the original process tensors have no common causally ordered extension. Thus
recursive causal constraints encode compatibility information that
channelization can discard. Related dependencies of compatibility on communication constraints have been studied for quantum instruments~\cite{buscemiUnifyingDifferentNotions2023}.

\begin{proposition}[Channel compatibility does not imply temporal compatibility]
There exist locally compatible process tensor marginals
\(\{\mathcal T_{J_1},\mathcal T_{J_2}\}\) that admit no common causally
ordered global process tensor, while their channelizations admit a common
global quantum channel.
\end{proposition}

Hence, compatibility of the channelized family is insufficient to establish temporal compatibility.

Consider a classical bit process embedded in qubit systems on the time set
\(K=\{0,1,2,3\}\). Let \(J_1:=\{0,2,3\}\) and
\(J_2:=\{1,2,3\}\). Define two marginal process tensors as follows.

Write \(P_r:=|r\rangle\langle r|\), with superscripts identifying the
tensor legs.

\begin{figure}[t]
    \centering
    \includegraphics[width=\columnwidth]{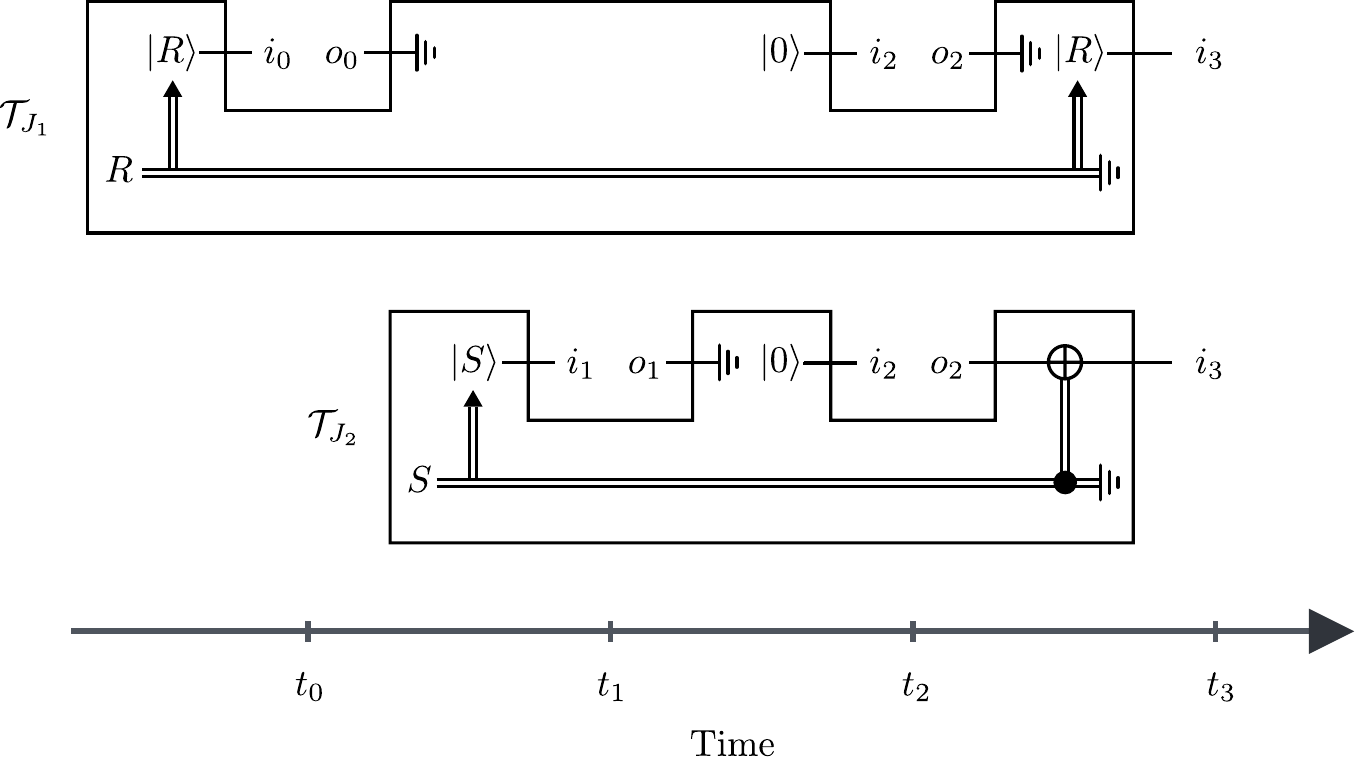}
    \caption{Temporal incompatibility hidden by channelization. The double horizontal lines carry the stored classical bits $R$ and $S$; terminal symbols denote discarded systems. For $J_1=\{0,2,3\}$, the process $\mathcal T_{J_1}$ samples $R$ uniformly, supplies $\lvert R\rangle$ on $i_0$ and $\lvert0\rangle$ on $i_2$, and returns $\lvert R\rangle$ on $i_3$. For $J_2=\{1,2,3\}$, the process $\mathcal T_{J_2}$ samples $S$ uniformly and supplies $\lvert S\rangle$ on $i_1$ and $\lvert0\rangle$ on $i_2$. After complete dephasing of the returned system $o_2$ in the computational basis, the displayed $S$-controlled $X$ maps each basis component $\lvert x\rangle$ to $\lvert S\oplus x\rangle$ on $i_3$. The marginals agree on their overlap $\{2,3\}$ and their channelizations admit a common quantum channel, but they have no common causally ordered process tensor extension.}
    \label{fig:irreducible-to-cmp}
\end{figure}

The marginal \(\mathcal T_{J_1}\) samples a uniformly random bit \(R\),
supplies \(|R\rangle\langle R|\) on \(i_0\), supplies
\(|0\rangle\langle0|\) on \(i_2\), and outputs
\(|R\rangle\langle R|\) on \(i_3\). Its joint classical state is
\begin{equation}
\frac{1}{2}\sum_{R=0}^1
|R\rangle\langle R|_{i_3}
\otimes
|0\rangle\langle 0|_{i_2}
\otimes
|R\rangle\langle R|_{i_0},
\end{equation}
and it discards the returned systems \(o_0\) and \(o_2\). Its complete
process-tensor Choi operator is
\begin{equation}
\Upsilon_{J_1}
=\frac12\sum_{r=0}^{1}
P_r^{i_3}\otimes I_{o_2}\otimes P_0^{i_2}
\otimes I_{o_0}\otimes P_r^{i_0}.
\end{equation}
The marginal
\(\mathcal T_{J_2}\) samples a uniformly random bit \(S\), supplies
\(|S\rangle\langle S|\) on \(i_1\), and supplies
\(|0\rangle\langle0|\) on \(i_2\). After dephasing the returned system
\(o_2\) in the computational basis and obtaining \(x\), it outputs
\(|S\oplus x\rangle\langle S\oplus x|\) on \(i_3\). The corresponding
joint classical state is
\begin{equation}
\frac{1}{2}\sum_{S=0}^1
|S\oplus x\rangle\langle S\oplus x|_{i_3}
\otimes
|0\rangle\langle 0|_{i_2}
\otimes
|S\rangle\langle S|_{i_1},
\end{equation}
and the process discards \(o_1\). Its complete process-tensor Choi
operator is
\begin{equation}
\Upsilon_{J_2}
=\frac12\sum_{s,x=0}^{1}
P_{s\oplus x}^{i_3}\otimes P_x^{o_2}
\otimes P_0^{i_2}\otimes I_{o_1}\otimes P_s^{i_1}.
\end{equation}
The preceding classical state expressions describe the prepared
computational-basis states; the complete process-tensor Choi operators also
include the control-output legs. In the second classical expression, \(x\)
labels a fixed computational-basis input on \(o_2\), whereas the complete
Choi operator sums over both \(s\) and \(x\). The identity factors on
discarded control-output legs implement the trace operation in the
unnormalized Choi convention.
The two marginal processes are illustrated in
Fig.~\ref{fig:irreducible-to-cmp}.

The two marginal process tensors have the same temporal marginal on
\(\{2,3\}\), whose complete Choi operator is
\begin{equation}
\Upsilon_{\{2,3\}}
=\frac{I_{i_3}}{2}\otimes I_{o_2}\otimes P_0^{i_2}.
\end{equation}
Equivalently, for every CP control \(\mathcal A_2\),
\begin{equation}
\mathcal T_{\{2,3\}}[\mathcal A_2]
=\operatorname{Tr}[\mathcal A_2(P_0)]\,\frac{I_{i_3}}{2}.
\end{equation}
For a trace-nonincreasing intervention outcome, the trace factor is its
success probability. For a CPTP control, this factor equals one. Thus the
prescribed marginals are locally compatible:
\begin{equation}
\mathcal M_{\{2,3\}}(\mathcal T_{J_1})
=
\mathcal M_{\{2,3\}}(\mathcal T_{J_2}).
\end{equation}

The causal conflict is visible in the correlations: the first marginal ties
the final bit to an earlier bit \(R\), whereas the second ties it to
\(S\oplus x\), with \(S\) supplied before the later control output \(x\)
is received. A common realization would have to reconcile these correlations
without allowing the later value \(x\) to influence earlier statistics.
Because the omitted slots are closed by identity channels, a rigorous proof
must retain the corresponding Choi contractions.

Suppose, for contradiction, that a common process tensor
\(\mathcal T_{3:0}\) exists, and let
\(\Upsilon:=J(\mathcal T_{3:0})\). Appendix~\ref{app:ptmp-not-cmp}
restricts \(\Upsilon\) to the subspaces labeled by the
computational-basis values \(c,x\in\{0,1\}\) at the corresponding legs,
giving four positive principal submatrices
\(\Upsilon_{cx}\succeq 0\), and constructs
positive operators \(W_{cx}\succeq 0\). The two marginal equalities,
with the omitted time-1 and time-0 slots contracted with
\(J(\mathcal I_1)\) and \(J(\mathcal I_0)\), respectively, together with
the final deterministic-comb causality condition
\begin{equation}
\operatorname{Tr}_{i_3}\Upsilon
=
\mathbb I_{o_2}\otimes\Upsilon_{2:0},
\end{equation}
imply
\begin{equation}
\sum_{c,x}
\operatorname{Tr}\!\left(
W_{cx}\Upsilon_{cx}
\right)
=
\frac{1-\sqrt{2}}{2}
<0.
\end{equation}
On the other hand, positivity of both
\(W_{cx}\) and \(\Upsilon_{cx}\) requires every term in the sum to be
nonnegative. This contradiction proves that no common causally ordered
global process tensor exists.

The obstruction disappears once the temporal ordering is forgotten. Regard
\(\mathcal T_{J_1}\) as a channel from \(o_2o_0\) to \(i_3i_2i_0\),
and \(\mathcal T_{J_2}\) as a channel from \(o_2o_1\) to
\(i_3i_2i_1\), both without causal constraints. Define a global quantum channel
\(\Phi:o_2o_1o_0\to i_3i_2i_1i_0\) by
\begin{equation}
\Phi(\rho_{o_2o_1o_0})
=
\sum_{x=0}^1
\operatorname{Tr}\!\left[
\left(
|x\rangle\langle x|_{o_2}\otimes \mathbb I_{o_1o_0}
\right)
\rho_{o_2o_1o_0}
\right]
\tau_x,
\end{equation}
where
\begin{equation}
\tau_x
:=
\frac{1}{2}\sum_{R=0}^1
|R\rangle\langle R|_{i_3}
\otimes
|0\rangle\langle 0|_{i_2}
\otimes
|R\oplus x\rangle\langle R\oplus x|_{i_1}
\otimes
|R\rangle\langle R|_{i_0}.
\end{equation}
Tracing out \(i_1\) gives the channelization of \(\mathcal T_{J_1}\), while
tracing out \(i_0\) gives the channelization of \(\mathcal T_{J_2}\) after
the relabeling \(S=R\oplus x\). Hence the channelized marginals are
compatible as ordinary channels, even though the original process tensor
marginals are incompatible.

A complete verification of validity and local compatibility, the explicit
Choi-level certificate of causal incompatibility, and the verification of
channel compatibility are given in
Appendix~\ref{app:ptmp-not-cmp}.

\section{Operational characterization by process discrimination}
\label{sec:operational-interpretation}

Distinguishing quantum processes is a fundamental task in quantum information. In a multi-time setting, an unknown process can be accessed through several temporally ordered interfaces, and information obtained at one time may be retained and used at later times. Its distinguishability may therefore depend on temporal correlations and memory that are absent from any single input-output interface~\cite{chiribellaMemoryEffectsQuantum2008,gutoskiMeasureDistanceQuantum2012,nakahiraGeneralizedQuantumProcess2021,Zambon:2024nnx}.

By a multi-time process-discrimination task, we mean a protocol in which a subtask label \(\alpha\) is selected, an unknown deterministic multi-time process \(\mathcal P_\alpha^{(i)}\) is drawn from a known ensemble with hidden label \(i\), the process is connected to the open slots of the corresponding marginal process tensor \(\mathcal T_\alpha\), and a measurement on the resulting final output state is used to guess \(i\). The task is specified by the subtask probabilities, the conditional process ensembles, and the corresponding final measurements, and its performance is quantified by the average success probability.

\begin{figure}[t]
    \centering
    \includegraphics[width=\columnwidth]{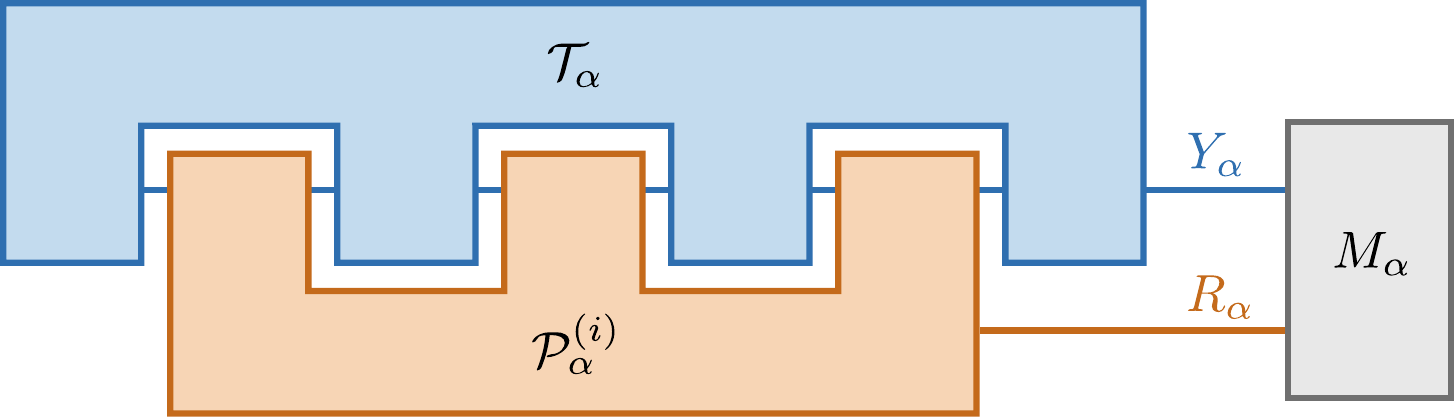}
    \caption{Multi-time process-discrimination task. A subtask label
    $\alpha$ is selected with probability $p_\alpha$ and revealed. Conditioned
    on $\alpha$, an unknown deterministic multi-time process
    $\mathcal P_\alpha^{(i)}$ is drawn with probability $q_\alpha^{(i)}$ and
    connected to the open slots of the marginal process tensor
    $\mathcal T_\alpha$. Their interaction produces the final state
    $\rho_{\alpha i}^{\mathcal T}$ on $Y_\alpha\otimes R_\alpha$, where
    $Y_\alpha$ is the final output system and $R_\alpha$ is the auxiliary
    readout system. The box labeled $M_\alpha$ represents the POVM
    $\{M_\alpha^{(j)}\}_j$ performed jointly on these systems. Its outcome
    $j$ is interpreted as the guess for the hidden label $i$, and the task
    succeeds when $j=i$.}
    \label{fig:discrimination-task}
\end{figure}

The task is illustrated in Fig.~\ref{fig:discrimination-task}.

Operationally, the label \(\alpha\) identifies which member
\(\mathcal T_\alpha\) of the family is available in a run. The label is
revealed before the run, and no global process tensor is assumed to exist.

Operational advantages in discrimination tasks are closely related to robustness and resourcefulness in general convex resource theories~\cite{takagiGeneralResourceTheories2019,uolaQuantificationQuantumDynamics2020,uolaQuantifyingQuantumResources2019}. Corresponding characterizations have been established for incompatibility of measurements, quantum channels, families of channel marginals, and quantum testers~\cite{carmeliWitnessingIncompatibilityQuantum2019,carmeliQuantumIncompatibilityWitnesses2019,skrzypczykAllSetsIncompatible2019,moriOperationalCharacterizationIncompatibility2020,Hsieh:2021ewp,sudarsananraginiHigherorderIncompatibilityImproves2024}. Related convex-optimization and robustness formulations have also been developed for quantum-process discrimination with restricted strategies~\cite{nakahiraQuantumProcessDiscrimination2021}. Here we establish an analogous characterization for incompatibility of process-tensor marginals: every incompatible family has an advantage over all compatible families in a suitably constructed multi-time process-discrimination task.

Specifically, we prove that a family is incompatible if and only if some
valid multi-time process-discrimination task gives it a strictly larger
success probability than every compatible family.

Let
\begin{equation}
    \boldsymbol{\mathcal T}
    :=
    \{\mathcal T_\alpha\}_{\alpha\in\Lambda}
\end{equation}
be a family of marginal process tensors, and let \(\mathfrak C\) denote the set of compatible families.

We consider a multi-time process-discrimination task with one subtask for each \(\alpha\in\Lambda\). With probability \(p_\alpha>0\), the subtask \(\alpha\) is selected, where
\begin{equation}
    \sum_{\alpha\in\Lambda}p_\alpha=1.
\end{equation}
Conditioned on \(\alpha\), an unknown multi-time process is drawn from the ensemble
\begin{equation}
    \left\{
        q_\alpha^{(i)},
        \mathcal P_\alpha^{(i)}
    \right\}_i,
    \qquad
    q_\alpha^{(i)}>0,
    \qquad
    \sum_i q_\alpha^{(i)}=1.
\end{equation}
Throughout this section, the processes to be discriminated are deterministic quantum combs~\cite{chiribellaQuantumCircuitArchitecture2008,chiribellaTheoreticalFrameworkQuantum2009}.

For each fixed \(\alpha\), the processes \(\{\mathcal P_\alpha^{(i)}\}_i\) have the same interactive input-output interfaces, matching the open slots of \(\mathcal T_\alpha\), and the same finite-dimensional auxiliary readout system \(R_\alpha\). This auxiliary system is available to the final measurement and may be one-dimensional. In the proof of Theorem~\ref{thm:discrimination-advantage}, it allows the completely positive multi-time control operations supplied by Theorem~\ref{thm:process-tensor-incompatibility-witness} to be embedded into deterministic processes.

Let \(Y_\alpha\) denote the final output system of \(\mathcal T_\alpha\). Connecting \(\mathcal P_\alpha^{(i)}\) to \(\mathcal T_\alpha\) produces the final output state
\begin{equation}
    \rho_{\alpha i}^{\mathcal T}
    :=
    \left(
        \mathcal T_\alpha
        \otimes
        \mathcal I_{R_\alpha}
    \right)
    \left[
        \mathcal P_\alpha^{(i)}
    \right]
\end{equation}
on \(Y_\alpha\otimes R_\alpha\). Since both \(\mathcal T_\alpha\) and \(\mathcal P_\alpha^{(i)}\) are deterministic,
\begin{equation}
    \operatorname{Tr}
    \rho_{\alpha i}^{\mathcal T}
    =
    1.
\end{equation}

A positive operator-valued measurement (POVM)
\begin{equation}
    \{M_\alpha^{(i)}\}_i,
    \qquad
    M_\alpha^{(i)}\succeq0,
    \qquad
    \sum_i M_\alpha^{(i)}
    =
    \mathbb I_{Y_\alpha\otimes R_\alpha},
\end{equation}
is performed on the final output state. Outcome \(i\) is interpreted as the guess that the supplied process was \(\mathcal P_\alpha^{(i)}\).

We denote the resulting discrimination task by
\begin{equation}
    D
    :=
    \left(
        \{p_\alpha\}_\alpha,
        \{
            q_\alpha^{(i)},
            \mathcal P_\alpha^{(i)}
        \}_{\alpha,i},
        \{
            M_\alpha^{(i)}
        \}_{\alpha,i}
    \right).
\end{equation}
Its success probability for the family \(\boldsymbol{\mathcal T}\) is
\begin{align}
    P(D,\boldsymbol{\mathcal T})
    &:=
    \sum_{\alpha\in\Lambda}
    p_\alpha
    \sum_i
    q_\alpha^{(i)}
    \operatorname{Tr}\left[
        M_\alpha^{(i)}
        \rho_{\alpha i}^{\mathcal T}
    \right]
    \nonumber\\
    &=
    \sum_{\alpha\in\Lambda}
    p_\alpha
    \sum_i
    q_\alpha^{(i)}
    \operatorname{Tr}\left[
        M_\alpha^{(i)}
        \left(
            \mathcal T_\alpha
            \otimes
            \mathcal I_{R_\alpha}
        \right)
        \left[
            \mathcal P_\alpha^{(i)}
        \right]
    \right].
\end{align}

For the same task, define the compatible benchmark as
\begin{equation}
    P_{\mathfrak C}(D)
    :=
    \max_{\boldsymbol{\Gamma}\in\mathfrak C}
    P(D,\boldsymbol{\Gamma}).
\end{equation}
The subtask probabilities, conditional process ensembles, and final measurements are fixed by \(D\); the maximization is only over compatible families.

\begin{theorem}[Advantage in multi-time process discrimination]
\label{thm:discrimination-advantage}
A family of marginal process tensors \(\boldsymbol{\mathcal T}\) is incompatible if and only if there exists a multi-time process-discrimination task \(D\) such that
\begin{equation}
    P(D,\boldsymbol{\mathcal T})
    >
    P_{\mathfrak C}(D).
\end{equation}
\end{theorem}

If \(\boldsymbol{\mathcal T}\in\mathfrak C\), then \(\boldsymbol{\mathcal T}\) is included in the maximization defining \(P_{\mathfrak C}(D)\) and therefore cannot exceed the compatible benchmark.

Conversely, Theorem~\ref{thm:process-tensor-incompatibility-witness} provides Hermitian final-output operators and completely positive multi-time control operations that separate \(\boldsymbol{\mathcal T}\) from every compatible family. Each such operation can be embedded into a deterministic multi-time process with a finite-dimensional auxiliary readout system while preserving the corresponding witness value. The resulting final output states are normalized, allowing the Hermitian operators to be shifted and rescaled into POVM effects. Completing the POVM with an additional process having sufficiently small prior probability preserves the strict separation. The detailed proof is given in Appendix~\ref{app:proof-operational-interpretation}.

Theorem~\ref{thm:discrimination-advantage} is an existence statement. It does not imply an advantage for every process ensemble. Rather, it shows that every incompatible family can be certified by some physically valid multi-time process-discrimination task.

\section{Conclusions}

We have formulated the process tensor marginal problem, which asks whether a family of multi-time processes specified on different subsets of times can arise as temporal marginals of a single global causal process. In contrast to state marginalization, temporal marginalization is operational and depends on the causal input-output structure of the process. We derived a semidefinite-programming characterization of global compatibility and showed that marginals supported on pairwise disjoint selected time sets are always compatible, even when their temporal intervals interleave. We also introduced an incompatibility robustness, derived its dual formulation, and obtained witness characterizations in terms of both Choi operators and multi-time control operations.

Our examples show that agreement on all pairwise overlaps is not sufficient for global compatibility. More importantly, the obstruction can be genuinely temporal: a family may fail to admit a common causally ordered process tensor while becoming compatible after the temporal ordering is forgotten and the marginals are treated as ordinary channels. We further showed that every incompatible family can be separated from all compatible families by a suitably constructed multi-time process-discrimination task. Temporal incompatibility is therefore both a failure of mathematical extension and an operationally detectable constraint imposed by memory and causal order.

These findings establish causal order as essential consistency information when combining process data obtained on different time sets. The compatibility SDP provides a test for the existence of a common causally ordered dynamical model, while the robustness determines the minimum mixing weight of another valid family needed to restore compatibility. The results also guide experimental design: since valid marginals on pairwise-disjoint selected time sets are always compatible, experiments seeking to reveal temporal incompatibility must probe overlapping time sets. For a specified incompatible family, the witness construction identifies a finite collection of multi-time controls and final measurements whose statistics can certify the conflict. These witnesses translate the mathematical obstruction into experimentally accessible tests of whether process descriptions reconstructed on different time sets can represent the same underlying dynamics.

A further experimentally relevant question is whether the available temporal marginals can be explained by memoryless dynamics at the sampled times. A Markovian description specifies the process through an initial state and a sequence of independent CPTP maps~\cite{Pollock:2018rut}, simplifying characterization and prediction. Testing whether this simpler description can reproduce the available data motivates imposing a Markovianity requirement on the global extension. The resulting Markovian process tensor marginal problem asks whether all prescribed marginals admit a common global Markovian process. If a family admits a global process tensor extension but no Markovian extension, then every global process consistent with the marginals necessarily contains memory. Resolving this problem would therefore allow partial multi-time information to rule out memoryless models without reconstructing the complete global process. Since exact Markovianity imposes factorization constraints on the process tensor that are generally nonconvex, suitable optimization methods and experimentally accessible certificates remain to be developed.

Beyond this structured-extension problem, several other directions remain
open. It would be valuable to characterize the overlap patterns among
selected time sets for which overlap consistency is sufficient for global
compatibility, and to identify minimal configurations that support genuinely
temporal compatibility obstructions. The separation established here also
raises the question of when channelization preserves compatibility and which
recursive causal constraints are lost under this reduction. Finally,
extending the framework to incomplete and noisy process-tensor data would
enable statistically robust compatibility tests in experimental settings.

\section*{Acknowledgments}
This work is supported by the Guangdong Basic and Applied Basic Research Foundation (Grant Nos.~2026A1515030035 and 2025A1515110223) and the Start-up Fund (Grant No.~G0101000274) from The Hong Kong University of Science and Technology (Guangzhou).

\bibliography{ref}

@article{Pollock:2018qgg,
  title = {Non-{{Markovian}} Quantum Processes: {{Complete}} Framework and Efficient Characterization},
  shorttitle = {Non-{{Markovian}} Quantum Processes},
  author = {Pollock, Felix A. and {Rodr{\'i}guez-Rosario}, C{\'e}sar and Frauenheim, Thomas and Paternostro, Mauro and Modi, Kavan},
  year = 2018,
  month = jan,
  journal = {Physical Review A},
  volume = {97},
  number = {1},
  pages = {012127},
  publisher = {American Physical Society},
  doi = {10.1103/PhysRevA.97.012127},
  urldate = {2026-03-02}
}

@book{watrousTheoryQuantumInformation2018a,
  title = {The {{Theory}} of {{Quantum Information}}},
  author = {Watrous, John},
  year = 2018,
  publisher = {Cambridge University Press},
  address = {Cambridge},
  doi = {10.1017/9781316848142},
  urldate = {2026-06-22},
  isbn = {978-1-107-18056-7}
}

@inproceedings{gutoskiGeneralTheoryQuantum2007,
  title = {Toward a General Theory of Quantum Games},
  booktitle = {Proceedings of the Thirty-Ninth Annual {{ACM}} Symposium on {{Theory}} of Computing},
  author = {Gutoski, Gus and Watrous, John},
  year = 2007,
  month = jun,
  eprint = {quant-ph/0611234},
  pages = {565--574},
  doi = {10.1145/1250790.1250873},
  urldate = {2026-06-22},
  archiveprefix = {arXiv}
}

@article{chiribellaQuantumCircuitArchitecture2008,
  title = {Quantum {{Circuit Architecture}}},
  author = {Chiribella, G. and D'Ariano, G. M. and Perinotti, P.},
  year = 2008,
  month = aug,
  journal = {Physical Review Letters},
  volume = {101},
  number = {6},
  pages = {060401},
  publisher = {American Physical Society},
  doi = {10.1103/PhysRevLett.101.060401},
  urldate = {2026-04-16}
}

@article{chiribellaTheoreticalFrameworkQuantum2009,
  title = {Theoretical Framework for Quantum Networks},
  author = {Chiribella, Giulio and D'Ariano, Giacomo Mauro and Perinotti, Paolo},
  year = 2009,
  month = aug,
  journal = {Physical Review A},
  volume = {80},
  number = {2},
  pages = {022339},
  publisher = {American Physical Society},
  doi = {10.1103/PhysRevA.80.022339},
  urldate = {2026-07-08}
}

@article{bravyiRequirementsCompatibilityLocal2004,
  title = {Requirements for Compatibility between Local and Multipartite Quantum States},
  author = {Bravyi, Sergey},
  year = 2004,
  month = jan,
  journal = {Quantum Information \& Computation},
  volume = {4},
  number = {1},
  pages = {12--26},
  issn = {1533-7146}
}

@article{klyachkoQuantumMarginalProblem2006,
  title = {Quantum Marginal Problem and {{N-representability}}},
  author = {Klyachko, Alexander A.},
  year = 2006,
  month = apr,
  journal = {Journal of Physics: Conference Series},
  volume = {36},
  number = {1},
  pages = {72},
  issn = {1742-6596},
  doi = {10.1088/1742-6596/36/1/014},
  urldate = {2026-07-08},
  langid = {english}
}

@article{Hsieh:2021ewp,
  title = {Quantum Channel Marginal Problem},
  author = {Hsieh, Chung-Yun and Lostaglio, Matteo and Ac{\'i}n, Antonio},
  year = 2022,
  month = mar,
  journal = {Physical Review Research},
  volume = {4},
  number = {1},
  pages = {013249},
  publisher = {American Physical Society},
  doi = {10.1103/PhysRevResearch.4.013249},
  urldate = {2026-03-02}
}

@article{whiteNonMarkovianQuantumProcess2022,
  title = {Non-{{Markovian Quantum Process Tomography}}},
  author = {White, G. A. L. and Pollock, F. A. and Hollenberg, L. C. L. and Modi, K. and Hill, C. D.},
  year = 2022,
  month = may,
  journal = {PRX Quantum},
  volume = {3},
  number = {2},
  pages = {020344},
  publisher = {American Physical Society},
  doi = {10.1103/PRXQuantum.3.020344},
  urldate = {2026-07-24}
}

@article{milzReconstructingNonMarkovianQuantum2018,
  title = {Reconstructing Non-{{Markovian}} Quantum Dynamics with Limited Control},
  author = {Milz, Simon and Pollock, Felix A. and Modi, Kavan},
  year = 2018,
  month = jul,
  journal = {Physical Review A},
  volume = {98},
  number = {1},
  pages = {012108},
  publisher = {American Physical Society},
  doi = {10.1103/PhysRevA.98.012108},
  urldate = {2026-07-24}
}

@article{whiteDemonstrationNonMarkovianProcess2020,
  title = {Demonstration of Non-{{Markovian}} Process Characterisation and Control on a Quantum Processor},
  author = {White, G. A. L. and Hill, C. D. and Pollock, F. A. and Hollenberg, L. C. L. and Modi, K.},
  year = 2020,
  month = dec,
  journal = {Nature Communications},
  volume = {11},
  number = {1},
  pages = {6301},
  publisher = {Nature Publishing Group},
  issn = {2041-1723},
  doi = {10.1038/s41467-020-20113-3},
  urldate = {2026-07-24},
  copyright = {2020 The Author(s)},
  langid = {english}
}

@article{berkExtractingQuantumDynamical2023,
  title = {Extracting Quantum Dynamical Resources: Consumption of Non-{{Markovianity}} for Noise Reduction},
  shorttitle = {Extracting Quantum Dynamical Resources},
  author = {Berk, Graeme D. and Milz, Simon and Pollock, Felix A. and Modi, Kavan},
  year = 2023,
  month = oct,
  journal = {npj Quantum Information},
  volume = {9},
  number = {1},
  pages = {104},
  publisher = {Nature Publishing Group},
  issn = {2056-6387},
  doi = {10.1038/s41534-023-00774-w},
  urldate = {2026-07-30},
  copyright = {2023 The Author(s)},
  langid = {english}
}

@article{chiribellaMemoryEffectsQuantum2008,
  title = {Memory {{Effects}} in {{Quantum Channel Discrimination}}},
  author = {Chiribella, Giulio and D'Ariano, Giacomo M. and Perinotti, Paolo},
  year = 2008,
  month = oct,
  journal = {Physical Review Letters},
  volume = {101},
  number = {18},
  pages = {180501},
  publisher = {American Physical Society},
  doi = {10.1103/PhysRevLett.101.180501},
  urldate = {2026-07-30}
}

@article{takagiGeneralResourceTheories2019,
  title = {General {{Resource Theories}} in {{Quantum Mechanics}} and {{Beyond}}: {{Operational Characterization}} via {{Discrimination Tasks}}},
  shorttitle = {General {{Resource Theories}} in {{Quantum Mechanics}} and {{Beyond}}},
  author = {Takagi, Ryuji and Regula, Bartosz},
  year = 2019,
  month = sep,
  journal = {Physical Review X},
  volume = {9},
  number = {3},
  pages = {031053},
  publisher = {American Physical Society},
  doi = {10.1103/PhysRevX.9.031053},
  urldate = {2026-07-30}
}

@article{hsiehResourceMarginalProblems2024,
  title = {Resource {{Marginal Problems}}},
  author = {Hsieh, Chung-Yun and Tabia, Gelo Noel M. and Yin, Yu-Chun and Liang, Yeong-Cherng},
  year = 2024,
  month = may,
  journal = {Quantum},
  volume = {8},
  pages = {1353},
  publisher = {Verein zur F\"orderung des Open Access Publizierens in den Quantenwissenschaften},
  doi = {10.22331/q-2024-05-22-1353},
  urldate = {2026-07-30},
  langid = {british}
}

@article{berkResourceTheoriesMultitime2021,
  title = {Resource Theories of Multi-Time Processes: {{A}} Window into Quantum Non-{{Markovianity}}},
  shorttitle = {Resource Theories of Multi-Time Processes},
  author = {Berk, Graeme D. and Garner, Andrew J. P. and Yadin, Benjamin and Modi, Kavan and Pollock, Felix A.},
  year = 2021,
  month = apr,
  journal = {Quantum},
  volume = {5},
  pages = {435},
  publisher = {Verein zur F\"orderung des Open Access Publizierens in den Quantenwissenschaften},
  doi = {10.22331/q-2021-04-20-435},
  urldate = {2026-07-30},
  langid = {british}
}

@article{carmeliWitnessingIncompatibilityQuantum2019,
  title = {Witnessing Incompatibility of Quantum Channels},
  author = {Carmeli, Claudio and Heinosaari, Teiko and Miyadera, Takayuki and Toigo, Alessandro},
  year = 2019,
  month = dec,
  journal = {Journal of Mathematical Physics},
  volume = {60},
  number = {12},
  pages = {122202},
  issn = {0022-2488},
  doi = {10.1063/1.5126496},
  urldate = {2026-08-01}
}

@article{carmeliQuantumIncompatibilityWitnesses2019,
  title = {Quantum {{Incompatibility Witnesses}}},
  author = {Carmeli, Claudio and Heinosaari, Teiko and Toigo, Alessandro},
  year = 2019,
  month = apr,
  journal = {Physical Review Letters},
  volume = {122},
  number = {13},
  pages = {130402},
  publisher = {American Physical Society},
  doi = {10.1103/PhysRevLett.122.130402},
  urldate = {2026-08-01}
}

@article{hallCompatibilitySubsystemStates2007,
  title = {Compatibility of Subsystem States and Convex Geometry},
  author = {Hall, William},
  year = 2007,
  month = mar,
  journal = {Physical Review A},
  volume = {75},
  number = {3},
  pages = {032102},
  publisher = {American Physical Society},
  doi = {10.1103/PhysRevA.75.032102},
  urldate = {2026-08-01}
}

@article{rossetResourceTheoryQuantum2018,
  title = {Resource {{Theory}} of {{Quantum Memories}} and {{Their Faithful Verification}} with {{Minimal Assumptions}}},
  author = {Rosset, Denis and Buscemi, Francesco and Liang, Yeong-Cherng},
  year = 2018,
  month = may,
  journal = {Physical Review X},
  volume = {8},
  number = {2},
  pages = {021033},
  publisher = {American Physical Society},
  doi = {10.1103/PhysRevX.8.021033},
  urldate = {2026-08-01}
}

@article{uolaQuantificationQuantumDynamics2020,
  title = {Quantification of Quantum Dynamics with Input-Output Games},
  author = {Uola, Roope and Kraft, Tristan and Abbott, Alastair A.},
  year = 2020,
  month = may,
  journal = {Physical Review A},
  volume = {101},
  number = {5},
  pages = {052306},
  publisher = {American Physical Society},
  doi = {10.1103/PhysRevA.101.052306},
  urldate = {2026-08-01}
}

@article{skrzypczykAllSetsIncompatible2019,
  title = {All {{Sets}} of {{Incompatible Measurements}} Give an {{Advantage}} in {{Quantum State Discrimination}}},
  author = {Skrzypczyk, Paul and {\v S}upi{\'c}, Ivan and Cavalcanti, Daniel},
  year = 2019,
  month = apr,
  journal = {Physical Review Letters},
  volume = {122},
  number = {13},
  pages = {130403},
  publisher = {American Physical Society},
  doi = {10.1103/PhysRevLett.122.130403},
  urldate = {2026-08-01}
}

@article{moriOperationalCharacterizationIncompatibility2020,
  title = {Operational Characterization of Incompatibility of Quantum Channels with Quantum State Discrimination},
  author = {Mori, Junki},
  year = 2020,
  month = mar,
  journal = {Physical Review A},
  volume = {101},
  number = {3},
  pages = {032331},
  publisher = {American Physical Society},
  doi = {10.1103/PhysRevA.101.032331},
  urldate = {2026-08-01}
}

@article{yuanUniversalOperationalBenchmarking2021,
  title = {Universal and Operational Benchmarking of Quantum Memories},
  author = {Yuan, Xiao and Liu, Yunchao and Zhao, Qi and Regula, Bartosz and Thompson, Jayne and Gu, Mile},
  year = 2021,
  month = jul,
  journal = {npj Quantum Information},
  volume = {7},
  number = {1},
  pages = {108},
  publisher = {Nature Publishing Group},
  issn = {2056-6387},
  doi = {10.1038/s41534-021-00444-9},
  urldate = {2026-08-01},
  copyright = {2021 The Author(s)},
  langid = {english}
}

@article{uolaQuantifyingQuantumResources2019,
  title = {Quantifying {{Quantum Resources}} with {{Conic Programming}}},
  author = {Uola, Roope and Kraft, Tristan and Shang, Jiangwei and Yu, Xiao-Dong and G{\"u}hne, Otfried},
  year = 2019,
  month = apr,
  journal = {Physical Review Letters},
  volume = {122},
  number = {13},
  pages = {130404},
  publisher = {American Physical Society},
  doi = {10.1103/PhysRevLett.122.130404},
  urldate = {2026-08-01}
}

@article{haapasaloRobustnessIncompatibilityQuantum2015,
  title = {Robustness of Incompatibility for Quantum Devices},
  author = {Haapasalo, Erkka},
  year = 2015,
  month = jun,
  journal = {Journal of Physics A: Mathematical and Theoretical},
  volume = {48},
  number = {25},
  pages = {255303},
  publisher = {IOP Publishing},
  issn = {1751-8121},
  doi = {10.1088/1751-8113/48/25/255303},
  urldate = {2026-08-01},
  langid = {english}
}

@article{haapasaloQuantumMarginalProblem2021,
  title = {Quantum Marginal Problem and Incompatibility},
  author = {Haapasalo, Erkka and Kraft, Tristan and Miklin, Nikolai and Uola, Roope},
  year = 2021,
  month = jun,
  journal = {Quantum},
  volume = {5},
  pages = {476},
  publisher = {Verein zur F\"orderung des Open Access Publizierens in den Quantenwissenschaften},
  doi = {10.22331/q-2021-06-15-476},
  urldate = {2026-07-27},
  langid = {british}
}

@article{sedlakIncompatibleMeasurementsQuantum2016,
  title = {Incompatible Measurements on Quantum Causal Networks},
  author = {Sedl{\'a}k, Michal and Reitzner, Daniel and Chiribella, Giulio and Ziman, M{\'a}rio},
  year = 2016,
  month = may,
  journal = {Physical Review A},
  volume = {93},
  number = {5},
  pages = {052323},
  publisher = {American Physical Society},
  doi = {10.1103/PhysRevA.93.052323},
  urldate = {2026-08-01}
}

@article{nakahiraGeneralizedQuantumProcess2021,
  title = {Generalized Quantum Process Discrimination Problems},
  author = {Nakahira, Kenji and Kato, Kentaro},
  year = 2021,
  month = jun,
  journal = {Physical Review A},
  volume = {103},
  number = {6},
  pages = {062606},
  publisher = {American Physical Society},
  doi = {10.1103/PhysRevA.103.062606},
  urldate = {2026-08-02}
}

@article{nakahiraQuantumProcessDiscrimination2021,
  title = {Quantum Process Discrimination with Restricted Strategies},
  author = {Nakahira, Kenji},
  year = 2021,
  month = dec,
  journal = {Physical Review A},
  volume = {104},
  number = {6},
  pages = {062609},
  publisher = {American Physical Society},
  doi = {10.1103/PhysRevA.104.062609},
  urldate = {2026-08-02}
}

@article{Zambon:2024nnx,
  title = {Process Tensor Distinguishability Measures},
  author = {Zambon, Guilherme},
  year = 2024,
  month = oct,
  journal = {Physical Review A},
  volume = {110},
  number = {4},
  pages = {042210},
  publisher = {American Physical Society},
  doi = {10.1103/PhysRevA.110.042210},
  urldate = {2026-03-02}
}

@article{jamiolkowskiLinearTransformationsWhich1972,
  title = {Linear Transformations Which Preserve Trace and Positive Semidefiniteness of Operators},
  author = {Jamio{\l}kowski, A.},
  year = 1972,
  month = dec,
  journal = {Reports on Mathematical Physics},
  volume = {3},
  number = {4},
  pages = {275--278},
  issn = {0034-4877},
  doi = {10.1016/0034-4877(72)90011-0},
  urldate = {2026-09-30}
}

@article{choiCompletelyPositiveLinear1975,
  title = {Completely Positive Linear Maps on Complex Matrices},
  author = {Choi, Man-Duen},
  year = 1975,
  month = jun,
  journal = {Linear Algebra and its Applications},
  volume = {10},
  number = {3},
  pages = {285--290},
  issn = {0024-3795},
  doi = {10.1016/0024-3795(75)90075-0},
  urldate = {2026-09-30}
}

@article{Pollock:2018rut,
  title = {Operational {{Markov Condition}} for {{Quantum Processes}}},
  author = {Pollock, Felix A. and {Rodr{\'i}guez-Rosario}, C{\'e}sar and Frauenheim, Thomas and Paternostro, Mauro and Modi, Kavan},
  year = 2018,
  month = jan,
  journal = {Physical Review Letters},
  volume = {120},
  number = {4},
  pages = {040405},
  publisher = {American Physical Society},
  doi = {10.1103/PhysRevLett.120.040405},
  urldate = {2026-03-02}
}

@article{Milz:2020usa,
  title = {Quantum {{Stochastic Processes}} and {{Quantum}} Non-{{Markovian Phenomena}}},
  author = {Milz, Simon and Modi, Kavan},
  year = 2021,
  month = jul,
  journal = {PRX Quantum},
  volume = {2},
  number = {3},
  pages = {030201},
  publisher = {American Physical Society},
  doi = {10.1103/PRXQuantum.2.030201},
  urldate = {2026-03-02}
}

@article{gutoskiMeasureDistanceQuantum2012,
  title = {On a Measure of Distance for Quantum Strategies},
  author = {Gutoski, Gus},
  year = 2012,
  month = mar,
  journal = {Journal of Mathematical Physics},
  volume = {53},
  number = {3},
  pages = {032202},
  issn = {0022-2488},
  doi = {10.1063/1.3693621},
  urldate = {2026-10-01}
}

@article{whiteWhatCanUnitary2025,
  title = {What Can Unitary Sequences Tell Us about Multi-Time Physics?},
  author = {White, Gregory A. L. and Pollock, Felix A. and Hollenberg, Lloyd C. L. and Hill, Charles D. and Modi, Kavan},
  year = 2025,
  month = apr,
  journal = {Quantum},
  volume = {9},
  pages = {1695},
  publisher = {Verein zur F\"orderung des Open Access Publizierens in den Quantenwissenschaften},
  doi = {10.22331/q-2025-04-08-1695},
  urldate = {2026-10-01},
  langid = {british}
}

@article{sudarsananraginiHigherorderIncompatibilityImproves2024,
  title = {Higher-Order Incompatibility Improves Distinguishability of Causal Quantum Networks},
  author = {Sudarsanan Ragini, Nidhin and Sazim, Sk},
  year = 2024,
  month = dec,
  journal = {New Journal of Physics},
  volume = {26},
  number = {12},
  pages = {123003},
  publisher = {IOP Publishing},
  issn = {1367-2630},
  doi = {10.1088/1367-2630/ad93f6},
  urldate = {2026-11-08},
  langid = {english}
}

@article{buscemiUnifyingDifferentNotions2023,
  title = {Unifying Different Notions of Quantum Incompatibility into a Strict Hierarchy of Resource Theories of Communication},
  author = {Buscemi, Francesco and Kobayashi, Kodai and Minagawa, Shintaro and Perinotti, Paolo and Tosini, Alessandro},
  year = 2023,
  month = jun,
  journal = {Quantum},
  volume = {7},
  pages = {1035},
  publisher = {Verein zur F\"orderung des Open Access Publizierens in den Quantenwissenschaften},
  doi = {10.22331/q-2023-06-07-1035},
  urldate = {2026-11-08},
  langid = {british}
}

\onecolumngrid
\appendix 
\section{Proof of Eq.~\eqref{eq:ptmp-sdp}}

\begin{lemma}
Let \(\{\Upsilon_\alpha\}_{\alpha\in\Lambda}\) be a collection of
Choi operators of prescribed marginal process tensors. The PTMP is feasible if and only if
the semidefinite feasibility problem in Eq.~\eqref{eq:ptmp-sdp} is
feasible.
\end{lemma}

\begin{proof}
If the PTMP is feasible, then there exists a global process tensor
\(\mathcal T_{k:0}\) satisfying
\(\mathcal M_\alpha(\mathcal T_{k:0})=\mathcal T_\alpha\) for all
\(\alpha\in\Lambda\). Its Choi operator
\(\Upsilon_{k:0}=J(\mathcal T_{k:0})\) is positive semidefinite and
satisfies the linear constraints in
Eq.~\eqref{eq:pt-causal-trace-constraints}. Taking the corresponding
intermediate Choi operators \(\Upsilon_{j:0}\), \(j=0,\ldots,k-1\), gives a
feasible point of Eq.~\eqref{eq:ptmp-sdp}.

Conversely, suppose Eq.~\eqref{eq:ptmp-sdp} is feasible. Then
\(\Upsilon_{k:0}\) is positive semidefinite and satisfies the linear constraints in Eq.~\eqref{eq:pt-causal-trace-constraints}.
These constraints are precisely the deterministic quantum-comb
normalization constraints, up to the relabeling of input and output spaces
and the one-dimensional initial input convention. In the notation of
Ref.~\cite{chiribellaQuantumCircuitArchitecture2008}, they correspond to
\begin{equation}
\operatorname{Tr}_{2n+1} R^{(n)}
=
I_{2n}\otimes R^{(n-1)},\qquad
R^{(N)}=R,\quad R^{(-1)}=1 .
\end{equation}
By Theorem~1 of Ref.~\cite{chiribellaQuantumCircuitArchitecture2008}, any positive operator
satisfying these recursive constraints is the Choi operator of a
deterministic quantum comb. By Corollary~3 of
Ref.~\cite{chiribellaTheoreticalFrameworkQuantum2009}, a deterministic \(N\)-comb is the
Choi-Jamio{\l}kowski operator of an \(N\)-partite memory channel. Hence
\(\Upsilon_{k:0}\) defines a valid causally ordered process tensor
\(\mathcal T_{k:0}\). The marginal constraints in
Eq.~\eqref{eq:ptmp-sdp} then imply that its marginals are precisely the
prescribed operators \(\{\Upsilon_\alpha\}_{\alpha\in\Lambda}\). Hence the
PTMP is feasible.
\end{proof}

\section{Proof of Eq.~\eqref{eq:ptmp-robustness-sdp}}

We prove that the robustness defined in Eq.~\eqref{eq:ptmp-robustness}
is given by the SDP in Eq.~\eqref{eq:ptmp-robustness-sdp}.

Throughout this appendix, whenever a family of process tensors
\(\boldsymbol{\mathcal X}
=\{\mathcal X_\alpha\}_{\alpha\in\Lambda}\) is given, we denote the Choi
operator of \(\mathcal X_\alpha\) by
\(\Upsilon_\alpha^{\mathcal X}:=J(\mathcal X_\alpha)\). Similarly, if a
global process tensor \(\mathcal X_{k:0}\) is given, we denote its Choi
operator by \(\Upsilon_{k:0}^{\mathcal X}:=J(\mathcal X_{k:0})\). In
particular, \(\Upsilon_\alpha^{\mathcal T}\) denotes the Choi operator
of the input marginal \(\mathcal T_\alpha\), which is denoted by
\(\Upsilon_\alpha\) in the main text.

By definition, \(R(\boldsymbol{\mathcal T})\) is the solution of
\begin{equation}
\begin{aligned}
\max_{\boldsymbol{\mathcal N},\boldsymbol{\mathcal L},\lambda}
\quad
& \lambda \\
\mathrm{subject\ to}
\quad
& \boldsymbol{\mathcal L}\in\mathfrak C, \\
& 0\le \lambda\le 1, \\
& \boldsymbol{\mathcal N}
=
\{\mathcal N_\alpha\}_{\alpha\in\Lambda}
\text{ is a family of valid marginal process tensors}, \\
& \lambda\boldsymbol{\mathcal T}
+
(1-\lambda)\boldsymbol{\mathcal N}
=
\boldsymbol{\mathcal L}.
\end{aligned}
\label{eq:robustness-auxiliary}
\end{equation}

Suppose first that
\((\boldsymbol{\mathcal N},\boldsymbol{\mathcal L},\lambda)\)
is feasible for Eq.~\eqref{eq:robustness-auxiliary}. Since
\(\boldsymbol{\mathcal L}\in\mathfrak C\), there exists a global process
tensor \(\mathcal L_{k:0}\) such that
\begin{equation}
\mathcal M_\alpha(\mathcal L_{k:0})
=
\mathcal L_\alpha,
\qquad
\forall\alpha\in\Lambda .
\label{eq:L-compatible-realization}
\end{equation}
In Choi form,
\begin{equation}
\mathcal M_\alpha(\Upsilon_{k:0}^{\mathcal L})
=
\Upsilon_\alpha^{\mathcal L},
\qquad
\forall\alpha\in\Lambda .
\end{equation}
Moreover, the last constraint in Eq.~\eqref{eq:robustness-auxiliary}
implies
\begin{equation}
\Upsilon_\alpha^{\mathcal L}
=
\lambda\Upsilon_\alpha^{\mathcal T}
+
(1-\lambda)\Upsilon_\alpha^{\mathcal N},
\qquad
\forall\alpha\in\Lambda .
\end{equation}
Since each \(\mathcal N_\alpha\) is valid, we have
\(\Upsilon_\alpha^{\mathcal N}\succeq0\). Therefore
\begin{equation}
\mathcal M_\alpha(\Upsilon_{k:0}^{\mathcal L})
=
\Upsilon_\alpha^{\mathcal L}
\succeq
\lambda\Upsilon_\alpha^{\mathcal T},
\qquad
\forall\alpha\in\Lambda .
\end{equation}
The Choi operator \(\Upsilon_{k:0}^{\mathcal L}\) of the global process
tensor \(\mathcal L_{k:0}\), together with its intermediate Choi
operators, satisfies the deterministic comb normalization constraints.
Hence
\((\lambda,\{\Upsilon_{j:0}^{\mathcal L}\}_{j=0}^{k})\)
is feasible for Eq.~\eqref{eq:ptmp-robustness-sdp}.

Conversely, suppose that
\((\lambda,\{\Upsilon_{j:0}^{\mathcal L}\}_{j=0}^{k})\)
is feasible for Eq.~\eqref{eq:ptmp-robustness-sdp}. By the same
deterministic-comb characterization used in the proof of
Eq.~\eqref{eq:ptmp-sdp}, \(\Upsilon_{k:0}^{\mathcal L}\) is the Choi
operator of a valid global process tensor \(\mathcal L_{k:0}\). Define
\begin{equation}
\Upsilon_\alpha^{\mathcal L}
:=
\mathcal M_\alpha(\Upsilon_{k:0}^{\mathcal L}),
\qquad
\forall\alpha\in\Lambda .
\end{equation}
Then
\(\boldsymbol{\mathcal L}:=\{\mathcal L_\alpha\}_{\alpha\in\Lambda}\),
with \(J(\mathcal L_\alpha)=\Upsilon_\alpha^{\mathcal L}\), belongs to
\(\mathfrak C\). The last constraint of
Eq.~\eqref{eq:ptmp-robustness-sdp} gives
\begin{equation}
\Upsilon_\alpha^{\mathcal L}
-
\lambda\Upsilon_\alpha^{\mathcal T}
\succeq0,
\qquad
\forall\alpha\in\Lambda .
\label{eq:residual-positive}
\end{equation}

Assume first that \(\lambda<1\), and define
\begin{equation}
\Upsilon_\alpha^{\mathcal N}
:=
\frac{
\Upsilon_\alpha^{\mathcal L}
-
\lambda\Upsilon_\alpha^{\mathcal T}
}{1-\lambda}.
\end{equation}
By Eq.~\eqref{eq:residual-positive},
\(\Upsilon_\alpha^{\mathcal N}\succeq0\). Moreover, \(\Upsilon_\alpha^{\mathcal L}\) and
\(\Upsilon_\alpha^{\mathcal T}\) are deterministic process tensor Choi
operators on the same selected time set \(\alpha\). Hence they satisfy
the same deterministic comb normalization constraints. Writing these
constraints compactly as
\begin{equation}
\mathcal A_\alpha(X)=b_\alpha ,
\end{equation}
we have
\begin{equation}
\mathcal A_\alpha(\Upsilon_\alpha^{\mathcal L})=b_\alpha,
\qquad
\mathcal A_\alpha(\Upsilon_\alpha^{\mathcal T})=b_\alpha .
\end{equation}
By linearity of \(\mathcal A_\alpha\),
\begin{equation}
\begin{aligned}
\mathcal A_\alpha(\Upsilon_\alpha^{\mathcal N})
&=
\frac{
\mathcal A_\alpha(\Upsilon_\alpha^{\mathcal L})
-
\lambda\mathcal A_\alpha(\Upsilon_\alpha^{\mathcal T})
}{1-\lambda} \\
&=
\frac{b_\alpha-\lambda b_\alpha}{1-\lambda}
=
b_\alpha .
\end{aligned}
\end{equation}
Therefore \(\Upsilon_\alpha^{\mathcal N}\) is positive and satisfies the
deterministic comb normalization constraints on \(\alpha\). By the
deterministic-comb characterization used in the proof of
Eq.~\eqref{eq:ptmp-sdp}, it is the Choi operator of a valid marginal
process tensor, which we denote by \(\mathcal N_\alpha\).

Thus
\begin{equation}
\Upsilon_\alpha^{\mathcal L}
=
\lambda\Upsilon_\alpha^{\mathcal T}
+
(1-\lambda)\Upsilon_\alpha^{\mathcal N},
\qquad
\forall\alpha\in\Lambda ,
\end{equation}
or equivalently,
\begin{equation}
\mathcal L_\alpha
=
\lambda\mathcal T_\alpha
+
(1-\lambda)\mathcal N_\alpha,
\qquad
\forall\alpha\in\Lambda .
\end{equation}
Since \(\boldsymbol{\mathcal L}\in\mathfrak C\), this gives a feasible
point of Eq.~\eqref{eq:robustness-auxiliary} with the same value of
\(\lambda\).

It remains to consider the boundary case \(\lambda=1\). Then
Eq.~\eqref{eq:residual-positive} becomes
\begin{equation}
\Upsilon_\alpha^{\mathcal L}
-
\Upsilon_\alpha^{\mathcal T}
\succeq0 .
\end{equation}
Since both operators are deterministic process tensor Choi operators on
the same selected time set, they have the same trace. Hence
\begin{equation}
\operatorname{Tr}
\bigl(
\Upsilon_\alpha^{\mathcal L}
-
\Upsilon_\alpha^{\mathcal T}
\bigr)
=
0 .
\end{equation}
A positive semidefinite operator with zero trace is zero, so
\begin{equation}
\Upsilon_\alpha^{\mathcal L}
=
\Upsilon_\alpha^{\mathcal T},
\qquad
\forall\alpha\in\Lambda .
\end{equation}
Therefore \(\boldsymbol{\mathcal T}\in\mathfrak C\). Taking
\(\boldsymbol{\mathcal L}=\boldsymbol{\mathcal T}\) and choosing any
valid family \(\boldsymbol{\mathcal N}\) gives a feasible point of
Eq.~\eqref{eq:robustness-auxiliary}.

We have shown that feasible points of Eq.~\eqref{eq:robustness-auxiliary}
and feasible points of Eq.~\eqref{eq:ptmp-robustness-sdp} can be
converted into one another without changing the value of \(\lambda\).
Therefore the two optimization problems have the same optimum, proving
Eq.~\eqref{eq:ptmp-robustness-sdp}.

\section{Proof of Eq.~\eqref{eq:ptmp-robustness-dual}}
\label{app:ptmp-robustness-dual}

In what follows, \(\mathcal{L}(\mathcal{H})\) denotes the set of linear
operators on a Hilbert space \(\mathcal{H}\). We let \(\mathbf{0}_x\)
denote the zero operator on the system
\begin{equation}
\mathcal{H}_{i_x}
\otimes
\left[
\bigotimes_{l=1}^{x}
\left(
\mathcal{H}_{o_{x-l}}
\otimes
\mathcal{H}_{i_{x-l}}
\right)
\right],
\end{equation}
and \(\mathbf{0}_x^{-}\) denote the zero operator on the reduced system
\begin{equation}
\bigotimes_{l=1}^{x}
\left(
\mathcal{H}_{o_{x-l}}
\otimes
\mathcal{H}_{i_{x-l}}
\right).
\end{equation}
For a finite set \(S=\{s_0,s_1,\dots,s_n\}\) satisfying
\(s_0<s_1<\cdots<s_n\), the symbol \(\widetilde{\mathbf{0}}_S\) denotes the
zero operator on
\begin{equation}
\mathcal{H}_{i_{s_n}}
\otimes
\left[
\bigotimes_{l=1}^{n}
\left(
\mathcal{H}_{o_{s_{n-l}}}
\otimes
\mathcal{H}_{i_{s_{n-l}}}
\right)
\right].
\end{equation}

Let
\begin{equation}
V :=
\left(
\bigoplus_{j=0}^{k}
\Upsilon_{j:0}
\right)
\oplus
\lambda
\end{equation}
and
\begin{equation}
A :=
\left(
\bigoplus_{j=0}^{k}
\mathbf{0}_{j}
\right)
\oplus
1,
\end{equation}
both in
\begin{equation}
\left[
\bigoplus_{j=0}^{k}
\mathcal{L}
\left(
\mathcal{H}_{i_j}
\otimes
\left[
\bigotimes_{l=1}^{j}
\left(
\mathcal{H}_{o_{j-l}}
\otimes
\mathcal{H}_{i_{j-l}}
\right)
\right]
\right)
\right]
\oplus
\mathbb{R},
\end{equation}
such that
\begin{equation}
\langle V,A\rangle
:=
\operatorname{Tr}(V^\dagger A)
=
\lambda.
\end{equation}

The feasible set can be characterized by defining the following functions:
\begin{equation}
\Phi
:=
\left(
\bigoplus_{j=1}^{k}
\Phi_j
\right)
\oplus
\Phi_0
\end{equation}
and
\begin{equation}
\Psi
:=
\left(
\bigoplus_{\alpha\in\Lambda}
\Psi_\alpha
\right)
\oplus
\Psi_0.
\end{equation}
For each \(j\in\{1,\dots,k\}\), we have
\begin{equation}
\Phi_j(V)
:=
\operatorname{Tr}_{i_j}(\Upsilon_{j:0})
-
\mathbb{I}_{o_{j-1}}
\otimes
\Upsilon_{j-1:0},
\end{equation}
and
\begin{equation}
\Phi_0(V)
:=
\operatorname{Tr}(\Upsilon_{0:0}).
\end{equation}
For each \(\alpha=\{a_0,a_1,\dots,a_n\}\in\Lambda\) satisfying
\(a_0<a_1<\cdots<a_n\leq k\), we have
\begin{equation}
\Psi_\alpha(V)
:=
\lambda\Upsilon_\alpha
-
\mathcal{M}_\alpha(\Upsilon_{k:0})
\end{equation}
and
\begin{equation}
\Psi_0(V)
:=
\lambda.
\end{equation}
Here
\begin{equation}
\begin{aligned}
\Phi_j
\colon
&
\left[
\bigoplus_{j=0}^{k}
\mathcal{L}
\left(
\mathcal{H}_{i_j}
\otimes
\left[
\bigotimes_{l=1}^{j}
\left(
\mathcal{H}_{o_{j-l}}
\otimes
\mathcal{H}_{i_{j-l}}
\right)
\right]
\right)
\right]
\oplus
\mathbb{R} \\
&\longrightarrow
\mathcal{L}
\left[
\bigotimes_{l=1}^{j}
\left(
\mathcal{H}_{o_{j-l}}
\otimes
\mathcal{H}_{i_{j-l}}
\right)
\right],
\end{aligned}
\end{equation}
\begin{equation}
\begin{aligned}
\Phi_0
\colon
&
\left[
\bigoplus_{j=0}^{k}
\mathcal{L}
\left(
\mathcal{H}_{i_j}
\otimes
\left[
\bigotimes_{l=1}^{j}
\left(
\mathcal{H}_{o_{j-l}}
\otimes
\mathcal{H}_{i_{j-l}}
\right)
\right]
\right)
\right]
\oplus
\mathbb{R} \\
&\longrightarrow
\mathbb{R},
\end{aligned}
\end{equation}
\begin{equation}
\begin{aligned}
\Psi_\alpha
\colon
&
\left[
\bigoplus_{j=0}^{k}
\mathcal{L}
\left(
\mathcal{H}_{i_j}
\otimes
\left[
\bigotimes_{l=1}^{j}
\left(
\mathcal{H}_{o_{j-l}}
\otimes
\mathcal{H}_{i_{j-l}}
\right)
\right]
\right)
\right]
\oplus
\mathbb{R} \\
&\longrightarrow
\mathcal{L}
\left(
\mathcal{H}_{i_{a_n}}
\otimes
\left[
\bigotimes_{l=1}^{n}
\left(
\mathcal{H}_{o_{a_{n-l}}}
\otimes
\mathcal{H}_{i_{a_{n-l}}}
\right)
\right]
\right),
\end{aligned}
\end{equation}
and
\begin{equation}
\begin{aligned}
\Psi_0
\colon
&
\left[
\bigoplus_{j=0}^{k}
\mathcal{L}
\left(
\mathcal{H}_{i_j}
\otimes
\left[
\bigotimes_{l=1}^{j}
\left(
\mathcal{H}_{o_{j-l}}
\otimes
\mathcal{H}_{i_{j-l}}
\right)
\right]
\right)
\right]
\oplus
\mathbb{R} \\
&\longrightarrow
\mathbb{R}.
\end{aligned}
\end{equation}
Also, we choose
\begin{equation}
B
:=
\left(
\bigoplus_{j=1}^{k}
\mathbf{0}_{j}^{-}
\right)
\oplus
1
\end{equation}
and
\begin{equation}
C
:=
\left(
\bigoplus_{\alpha\in\Lambda}
\widetilde{\mathbf{0}}_\alpha
\right)
\oplus
1.
\end{equation}
One can rewrite the SDP as
\begin{equation}
\begin{aligned}
\max_{V}\quad
& \langle V,A\rangle \\
\mathrm{subject\ to}\quad
& \Phi(V)=B, \\
& \Psi(V)\preceq C, \\
& V\succeq 0.
\end{aligned}
\end{equation}
The corresponding dual problem is
\begin{equation}
\begin{aligned}
\min_{H,Z}\quad
& \langle H,B\rangle+\langle Z,C\rangle \\
\mathrm{subject\ to}\quad
& \Phi^\dagger(H)+\Psi^\dagger(Z)\succeq A, \\
& H^\dagger=H, \\
& Z\succeq 0.
\end{aligned}
\end{equation}
Here
\begin{equation}
H
=
\left(
\bigoplus_{j=1}^{k}
H_j
\right)
\oplus
h
\in
\left[
\bigoplus_{j=1}^{k}
\mathcal{L}
\left(
\bigotimes_{l=1}^{j}
\left(
\mathcal{H}_{o_{j-l}}
\otimes
\mathcal{H}_{i_{j-l}}
\right)
\right)
\right]
\oplus
\mathbb{R}
\end{equation}
and
\begin{equation}
Z
=
\left(
\bigoplus_{\alpha\in\Lambda}
Z_\alpha
\right)
\oplus
z
\in
\left[
\bigoplus_{\alpha\in\Lambda}
\mathcal{L}
\left(
\mathcal{H}_{i_{a_n}}
\otimes
\left[
\bigotimes_{l=1}^{n}
\left(
\mathcal{H}_{o_{a_{n-l}}}
\otimes
\mathcal{H}_{i_{a_{n-l}}}
\right)
\right]
\right)
\right]
\oplus
\mathbb{R}.
\end{equation}

Since $\Phi$ and $\Psi$ are direct sums of their component maps,
their adjoints decompose componentwise. Namely,
\begin{equation}
\begin{aligned}
\Phi^\dagger(H)
&=
\Phi_0^\dagger(h)
+
\sum_{j=1}^{k}\Phi_j^\dagger(H_j),
\\
\Psi^\dagger(Z)
&=
\Psi_0^\dagger(z)
+
\sum_{\alpha\in\Lambda}
\Psi_\alpha^\dagger(Z_\alpha).
\end{aligned}
\end{equation}
It therefore suffices to compute the adjoints of
$\Phi_0$, $\Phi_j$, $\Psi_0$, and $\Psi_\alpha$ separately.

Direct computation shows
\begin{equation}
\begin{aligned}
\langle \Phi_0^\dagger(h),V\rangle
&=
\langle h,\Phi_0(V)\rangle \\
&=
\langle h,\operatorname{Tr}(\Upsilon_{0:0})\rangle \\
&=
h\operatorname{Tr}(\Upsilon_{0:0}) \\
&=
\langle h\mathbb{I}_{i_0},\Upsilon_{0:0}\rangle,
\end{aligned}
\end{equation}
which means
\begin{equation}
\Phi_0^\dagger(h)
=
h\mathbb{I}_{i_0}
\oplus
\left(
\bigoplus_{j=1}^{k}
\mathbf{0}_j
\right)
\oplus
0.
\end{equation}
For a given \(j\in\{1,\dots,k\}\), we have
\begin{equation}
\begin{aligned}
\langle \Phi_j^\dagger(H_j),V\rangle
&=
\langle H_j,\Phi_j(V)\rangle \\
&=
\left\langle
H_j,
\operatorname{Tr}_{i_j}(\Upsilon_{j:0})
-
\mathbb{I}_{o_{j-1}}\otimes\Upsilon_{j-1:0}
\right\rangle \\
&=
\operatorname{Tr}\!\left[
H_j
\operatorname{Tr}_{i_j}(\Upsilon_{j:0})
\right]
-
\operatorname{Tr}\!\left[
H_j
\left(
\mathbb{I}_{o_{j-1}}
\otimes
\Upsilon_{j-1:0}
\right)
\right] \\
&=
\operatorname{Tr}\!\left[
\left(
\mathbb{I}_{i_j}
\otimes
H_j
\right)
\Upsilon_{j:0}
\right]
-
\operatorname{Tr}\!\left[
\Upsilon_{j-1:0}
\operatorname{Tr}_{o_{j-1}}(H_j)
\right] \\
&=
\left\langle
\left(
\bigoplus_{l=0}^{j-2}
\mathbf{0}_l
\right)
\oplus
\left[
-
\operatorname{Tr}_{o_{j-1}}(H_j)
\right]
\oplus
\left(
\mathbb{I}_{i_j}
\otimes
H_j
\right)
\oplus
\left(
\bigoplus_{l=j+1}^{k}
\mathbf{0}_l
\right)
\oplus
0,
V
\right\rangle,
\end{aligned}
\end{equation}
which implies
\begin{equation}
\Phi_j^\dagger(H_j)
=
\left(
\bigoplus_{l=0}^{j-2}
\mathbf{0}_l
\right)
\oplus
\left[
-
\operatorname{Tr}_{o_{j-1}}(H_j)
\right]
\oplus
\left(
\mathbb{I}_{i_j}
\otimes
H_j
\right)
\oplus
\left(
\bigoplus_{l=j+1}^{k}
\mathbf{0}_l
\right)
\oplus
0.
\end{equation}

Also, we have
\begin{equation}
\begin{aligned}
\langle \Psi_0^\dagger(z),V\rangle
&=
\langle z,\Psi_0(V)\rangle \\
&=
\langle z,\lambda\rangle \\
&=
z\lambda,
\end{aligned}
\end{equation}
which means
\begin{equation}
\Psi_0^\dagger(z)
=
\left(
\bigoplus_{j=0}^{k}
\mathbf{0}_j
\right)
\oplus
z.
\end{equation}
For a given \(\alpha=\{a_0,a_1,\dots,a_n\}\in\Lambda\) satisfying
\(a_0<a_1<\cdots<a_n\leq k\), we have
\begin{equation}
\begin{aligned}
\langle \Psi_\alpha^\dagger(Z_\alpha),V\rangle
&=
\langle Z_\alpha,\Psi_\alpha(V)\rangle \\
&=
\left\langle
Z_\alpha,
\lambda\Upsilon_\alpha
-
\mathcal{M}_\alpha(\Upsilon_{k:0})
\right\rangle \\
&=
\lambda
\operatorname{Tr}\!\left(
Z_\alpha\Upsilon_\alpha
\right)
-
\left\langle
\mathcal{M}_\alpha^\dagger(Z_\alpha),
\Upsilon_{k:0}
\right\rangle \\
&=
\left\langle
\left(
\bigoplus_{j=0}^{k-1}
\mathbf{0}_j
\right)
\oplus
\left[
-
\mathcal{M}_\alpha^\dagger(Z_\alpha)
\right]
\oplus
\operatorname{Tr}\!\left(
Z_\alpha\Upsilon_\alpha
\right),
V
\right\rangle,
\end{aligned}
\end{equation}
which implies
\begin{equation}
\Psi_\alpha^\dagger(Z_\alpha)
=
\left(
\bigoplus_{j=0}^{k-1}
\mathbf{0}_j
\right)
\oplus
\left[
-
\mathcal{M}_\alpha^\dagger(Z_\alpha)
\right]
\oplus
\operatorname{Tr}\!\left(
Z_\alpha\Upsilon_\alpha
\right).
\end{equation}

We have \(\langle H,B\rangle=h\) and \(\langle Z,C\rangle=z\). Direct
computation shows
\begin{equation}
\begin{aligned}
\Phi^\dagger(H)
&=
h\mathbb{I}_{i_0}
\oplus
\left(
\bigoplus_{j=1}^{k}
\mathbf{0}_j
\right)
\oplus
0 \\
&\quad
+
\sum_{j=1}^{k}
\left[
\left(
\bigoplus_{l=0}^{j-2}
\mathbf{0}_l
\right)
\oplus
\left[
-
\operatorname{Tr}_{o_{j-1}}(H_j)
\right]
\oplus
\left(
\mathbb{I}_{i_j}
\otimes
H_j
\right)
\oplus
\left(
\bigoplus_{l=j+1}^{k}
\mathbf{0}_l
\right)
\oplus
0
\right] \\
&=
\left(
h\mathbb{I}_{i_0}
-
\operatorname{Tr}_{o_0}(H_1)
\right)
\oplus
\left[
\bigoplus_{j=1}^{k-1}
\left(
\mathbb{I}_{i_j}
\otimes
H_j
-
\operatorname{Tr}_{o_j}(H_{j+1})
\right)
\right] \\
&\quad
\oplus
\left(
\mathbb{I}_{i_k}
\otimes
H_k
\right)
\oplus
0
\end{aligned}
\end{equation}
and
\begin{equation}
\begin{aligned}
\Psi^\dagger(Z)
&=
\left(
\bigoplus_{j=0}^{k}
\mathbf{0}_j
\right)
\oplus
z \\
&\quad
+
\sum_{\alpha\in\Lambda}
\left[
\left(
\bigoplus_{j=0}^{k-1}
\mathbf{0}_j
\right)
\oplus
\left[
-
\mathcal{M}_\alpha^\dagger(Z_\alpha)
\right]
\oplus
\operatorname{Tr}\!\left(
Z_\alpha\Upsilon_\alpha
\right)
\right] \\
&=
\left(
\bigoplus_{j=0}^{k-1}
\mathbf{0}_j
\right)
\oplus
\left[
-
\sum_{\alpha\in\Lambda}
\mathcal{M}_\alpha^\dagger(Z_\alpha)
\right]
\oplus
\left[
z
+
\sum_{\alpha\in\Lambda}
\operatorname{Tr}\!\left(
Z_\alpha\Upsilon_\alpha
\right)
\right].
\end{aligned}
\end{equation}
Thus
\begin{equation}
\begin{aligned}
\Phi^\dagger(H)+\Psi^\dagger(Z)
&=
\left(
h\mathbb{I}_{i_0}
-
\operatorname{Tr}_{o_0}(H_1)
\right)
\oplus
\left[
\bigoplus_{j=1}^{k-1}
\left(
\mathbb{I}_{i_j}
\otimes
H_j
-
\operatorname{Tr}_{o_j}(H_{j+1})
\right)
\right] \\
&\quad
\oplus
\left(
\mathbb{I}_{i_k}
\otimes
H_k
-
\sum_{\alpha\in\Lambda}
\mathcal{M}_\alpha^\dagger(Z_\alpha)
\right) \\
&\quad
\oplus
\left[
z
+
\sum_{\alpha\in\Lambda}
\operatorname{Tr}\!\left(
Z_\alpha\Upsilon_\alpha
\right)
\right].
\end{aligned}
\end{equation}
Combining everything, we have
\begin{equation}
\begin{aligned}
\min_{h,z,\{H_j\},\{Z_\alpha\}}
\quad
& h+z \\[1mm]
\mathrm{subject\ to}
\quad
& h\mathbb{I}_{i_0}
-
\operatorname{Tr}_{o_0}(H_1)
\succeq 0, \\[1mm]
& \mathbb{I}_{i_j}\otimes H_j
-
\operatorname{Tr}_{o_j}(H_{j+1})
\succeq 0,
&& j=1,\ldots,k-1, \\[1mm]
& \mathbb{I}_{i_k}\otimes H_k
-
\sum_{\alpha\in\Lambda}
\mathcal{M}_\alpha^\dagger(Z_\alpha)
\succeq 0, \\[1mm]
& z
+
\sum_{\alpha\in\Lambda}
\operatorname{Tr}\!\left(
Z_\alpha\Upsilon_\alpha
\right)
\geq 1, \\[1mm]
& H_j^\dagger=H_j,
&& j=1,\ldots,k, \\[1mm]
& h\in\mathbb{R}, \\[1mm]
& Z_\alpha\succeq 0,
&& \forall\alpha\in\Lambda, \\[1mm]
& z\geq 0.
\end{aligned}
\end{equation}

Finally, strong duality holds. The primal problem is feasible by taking
\(\lambda=0\) and choosing any valid normalized process tensor together with
its causal prefixes. For example, one may choose the recursively defined
depolarizing comb
\begin{equation}
\Upsilon_{0:0}
=
\frac{\mathbb{I}_{i_0}}{d_{i_0}},
\qquad
\Upsilon_{j:0}
=
\frac{\mathbb{I}_{i_j}}{d_{i_j}}
\otimes
\mathbb{I}_{o_{j-1}}
\otimes
\Upsilon_{j-1:0}.
\end{equation}
This satisfies
\begin{equation}
\operatorname{Tr}_{i_j}(\Upsilon_{j:0})
=
\mathbb{I}_{o_{j-1}}
\otimes
\Upsilon_{j-1:0}.
\end{equation}
Moreover, the primal objective is finite because \(0\leq\lambda\leq1\).
The dual is strictly feasible: choose \(Z_\alpha\succ0\) for all
\(\alpha\), then choose \(H_k,H_{k-1},\ldots,H_1\) recursively as
sufficiently large scalar multiples of the identity so that all operator
inequalities are strict, and finally choose \(h\in\mathbb{R}\) and \(z>0\)
sufficiently large so that the first and scalar constraints are strict.
Hence, by Slater's condition for finite-dimensional semidefinite programs~\cite{watrousTheoryQuantumInformation2018a},
strong duality holds.

\section{Proof of Eq.~\eqref{eq:choi-incompatibility-witness}}
\label{app:choi-incompatibility-witness}

The witness inequality immediately implies incompatibility, since a compatible
family is included in the maximization over \(\mathfrak C\).

Conversely, suppose \(\boldsymbol{\mathcal T}\) is incompatible. Then
\(R(\boldsymbol{\mathcal T})<1\). By strong duality, there is a feasible
tuple of Eq.~\eqref{eq:ptmp-robustness-dual} with \(h+z<1\).

Write \(\mathbf Z:=\{Z_\alpha\}_{\alpha\in\Lambda}\). For every
compatible family \(\boldsymbol{\Gamma}\), the recursive comb constraints
and the first three dual inequalities give
\[
\langle\mathbf Z,\boldsymbol{\Upsilon}^{\Gamma}\rangle\leq h.
\]
Indeed, choose a global extension with Choi operator
\(\Upsilon_{k:0}^{\Gamma}\) and let \(\Upsilon_{j:0}^{\Gamma}\) denote its
causal prefixes. Their recursive normalization equalities are
\[
\operatorname{Tr}_{i_j}(\Upsilon_{j:0}^{\Gamma})
=
\mathbb I_{o_{j-1}}\otimes\Upsilon_{j-1:0}^{\Gamma},
\qquad j=1,\ldots,k,
\]
and \(\operatorname{Tr}(\Upsilon_{0:0}^{\Gamma})=1\). Applying these
equalities between successive dual inequalities gives
\begin{equation}
\begin{aligned}
h
&=h\operatorname{Tr}(\Upsilon_{0:0}^{\Gamma})
\geq
\operatorname{Tr}\!\left[
    \operatorname{Tr}_{o_0}(H_1)\Upsilon_{0:0}^{\Gamma}
\right]
\\
&=\operatorname{Tr}\!\left[
    H_1(\mathbb I_{o_0}\otimes\Upsilon_{0:0}^{\Gamma})
\right]
=\operatorname{Tr}\!\left[
    (\mathbb I_{i_1}\otimes H_1)\Upsilon_{1:0}^{\Gamma}
\right]
\\
&\geq\cdots\geq
\operatorname{Tr}\!\left[
    (\mathbb I_{i_k}\otimes H_k)\Upsilon_{k:0}^{\Gamma}
\right]
\\
&\geq
\operatorname{Tr}\!\left[
    \left(\sum_{\alpha\in\Lambda}
    \mathcal M_\alpha^\dagger(Z_\alpha)\right)
    \Upsilon_{k:0}^{\Gamma}
\right]
=\sum_{\alpha\in\Lambda}
\operatorname{Tr}\!\left[
    Z_\alpha\mathcal M_\alpha(\Upsilon_{k:0}^{\Gamma})
\right]
\\
&=\sum_{\alpha\in\Lambda}
\operatorname{Tr}\!\left[
    Z_\alpha\Upsilon_\alpha^{\Gamma}
\right]
=\langle\mathbf Z,\boldsymbol{\Upsilon}^{\Gamma}\rangle.
\end{aligned}
\end{equation}

The scalar dual constraint gives
\[
\langle\mathbf Z,\boldsymbol{\Upsilon}^{\mathcal T}\rangle
\geq 1-z>h.
\]
Therefore \(\langle\mathbf Z,\boldsymbol{\Upsilon}^{\mathcal T}\rangle\)
exceeds the maximum compatible score. Setting \(W_\alpha:=Z_\alpha\) for
every \(\alpha\) proves the assertion.

\section{Proof of Theorem~\ref{thm:process-tensor-incompatibility-witness}}
\label{app:process-tensor-incompatibility-witness}

First, suppose that there exist finitely many Hermitian operators
\(F_\alpha^{(i)}=(F_\alpha^{(i)})^\dagger\) and finitely many completely
positive multi-time control operations \(\mathbf A_\alpha^{(i)}\) such that
\begin{equation}
\sum_{\alpha\in\Lambda}
\sum_i
\operatorname{Tr}\!\left[
F_\alpha^{(i)}
\,
\mathcal T_\alpha
\!\left[
\mathbf A_\alpha^{(i)}
\right]
\right]
>
\max_{\boldsymbol{\Gamma}\in\mathfrak C}
\sum_{\alpha\in\Lambda}
\sum_i
\operatorname{Tr}\!\left[
F_\alpha^{(i)}
\,
\Gamma_\alpha
\!\left[
\mathbf A_\alpha^{(i)}
\right]
\right].
\end{equation}
If \(\boldsymbol{\mathcal T}\in\mathfrak C\), then
\(\boldsymbol{\mathcal T}\) is one of the feasible process tensor families
in the maximization over \(\mathfrak C\). Hence the right-hand side is at
least
\begin{equation}
\sum_{\alpha\in\Lambda}
\sum_i
\operatorname{Tr}\!\left[
F_\alpha^{(i)}
\,
\mathcal T_\alpha
\!\left[
\mathbf A_\alpha^{(i)}
\right]
\right],
\end{equation}
which contradicts the strict inequality. Therefore,
\(\boldsymbol{\mathcal T}\notin\mathfrak C\).

Conversely, assume that \(\boldsymbol{\mathcal T}\notin\mathfrak C\). By
the Choi witness result, there exists a family of positive operators
\(\boldsymbol W:=\{W_\alpha\}_{\alpha\in\Lambda}\), with
\(W_\alpha\succeq0\), such that
\begin{equation}
\langle \boldsymbol W,\boldsymbol{\Upsilon}^{\mathcal T}\rangle
>
\max_{\boldsymbol{\Gamma}\in\mathfrak C}
\langle \boldsymbol W,\boldsymbol{\Upsilon}^{\Gamma}\rangle .
\end{equation}

Fix \(\alpha\in\Lambda\). By the spanning result for multi-time control
operations in the Choi representation~\cite{Pollock:2018qgg}, there exists
a finite family of completely positive multi-time control operations
\(\{\mathbf A_\alpha^{(r)}\}_r\) such that the corresponding Choi operators
\(\{J(\mathbf A_\alpha^{(r)})\}_r\) span the Hermitian operator space on the
control-slot Hilbert space of the marginal process tensor
\(\mathcal T_\alpha\). Since transposition is a linear bijection, the
transposed operators \(\{J(\mathbf A_\alpha^{(r)})^T\}_r\) also span that
Hermitian operator space. Let \(\{G_\alpha^{(s)}\}_s\) be a finite Hermitian
basis of the final output Hilbert space of \(\mathcal T_\alpha\). Then the
operators \(G_\alpha^{(s)}\otimes J(\mathbf A_\alpha^{(r)})^T\) span the
Hermitian operator space on the Choi space of \(\mathcal T_\alpha\). Hence,
for each \(W_\alpha\), there exist real coefficients
\(c_{\alpha,s,r}\in\mathbb R\) such that
\begin{equation}
W_\alpha
=
\sum_{s,r}
c_{\alpha,s,r}
G_\alpha^{(s)}
\otimes
J(\mathbf A_\alpha^{(r)})^T.
\end{equation}
Define \(F_\alpha^{(s,r)}:=c_{\alpha,s,r}G_\alpha^{(s)}\). Then each
\(F_\alpha^{(s,r)}\) is Hermitian and
\begin{equation}
W_\alpha
=
\sum_{s,r}
F_\alpha^{(s,r)}
\otimes
J(\mathbf A_\alpha^{(r)})^T.
\end{equation}

By the Choi action formula for process tensors, for any marginal process
tensor \(\Theta_\alpha\) with Choi operator \(\Upsilon_\alpha^\Theta\), we
have
\begin{equation}
\Theta_\alpha[\mathbf A_\alpha]
=
\operatorname{Tr}_{\mathrm{ctrl}}
\left[
\Upsilon_\alpha^\Theta
\left(
\mathbb I
\otimes
J(\mathbf A_\alpha)^T
\right)
\right],
\end{equation}
where the partial trace is over the control-slot Choi legs. Therefore, for
every Hermitian operator \(F_\alpha\) on the final output Hilbert space,
\begin{equation}
\operatorname{Tr}\!\left[
F_\alpha
\Theta_\alpha[\mathbf A_\alpha]
\right]
=
\operatorname{Tr}\!\left[
\left(
F_\alpha
\otimes
J(\mathbf A_\alpha)^T
\right)
\Upsilon_\alpha^\Theta
\right].
\end{equation}
Applying this identity to the above decomposition of \(W_\alpha\), we obtain
\begin{equation}
\operatorname{Tr}\!\left[
W_\alpha
\Upsilon_\alpha^\Theta
\right]
=
\sum_{s,r}
\operatorname{Tr}\!\left[
F_\alpha^{(s,r)}
\,
\Theta_\alpha
\!\left[
\mathbf A_\alpha^{(r)}
\right]
\right].
\end{equation}

Taking \(\Theta_\alpha=\mathcal T_\alpha\) and
\(\Theta_\alpha=\Gamma_\alpha\), and substituting these identities into the
Choi witness inequality, gives
\begin{equation}
\sum_{\alpha\in\Lambda}
\sum_{s,r}
\operatorname{Tr}\!\left[
F_\alpha^{(s,r)}
\,
\mathcal T_\alpha
\!\left[
\mathbf A_\alpha^{(r)}
\right]
\right]
>
\max_{\boldsymbol{\Gamma}\in\mathfrak C}
\sum_{\alpha\in\Lambda}
\sum_{s,r}
\operatorname{Tr}\!\left[
F_\alpha^{(s,r)}
\,
\Gamma_\alpha
\!\left[
\mathbf A_\alpha^{(r)}
\right]
\right].
\end{equation}
Relabelling the pair \((s,r)\) as a single finite index \(i\) yields
\begin{equation}
\sum_{\alpha\in\Lambda}
\sum_i
\operatorname{Tr}\!\left[
F_\alpha^{(i)}
\,
\mathcal T_\alpha
\!\left[
\mathbf A_\alpha^{(i)}
\right]
\right]
>
\max_{\boldsymbol{\Gamma}\in\mathfrak C}
\sum_{\alpha\in\Lambda}
\sum_i
\operatorname{Tr}\!\left[
F_\alpha^{(i)}
\,
\Gamma_\alpha
\!\left[
\mathbf A_\alpha^{(i)}
\right]
\right].
\end{equation}
This completes the proof.

\section{Example of incompatibility}
\label{app:locally-compatible-incompatible}

We now give an example showing that local compatibility of process tensor
marginals does not imply global compatibility. Consider a qubit process on
the time set \(K=\{0,1,2\}\). Following the subsystem ordering convention
introduced above, a global two-step process tensor \(\mathcal T_{2:0}\) has
open Hilbert space
\begin{equation}
\mathcal H_{2:0}
=
\mathcal H_{i_2}
\otimes
\mathcal H_{o_1}
\otimes
\mathcal H_{i_1}
\otimes
\mathcal H_{o_0}
\otimes
\mathcal H_{i_0}.
\end{equation}
Let \(\rho_0=\mathbb I/2\) be the initial state, and let \(\mathcal X\)
denote the bit-flip channel, \(\mathcal X(\rho)=X\rho X\), where \(X\) is
the Pauli bit-flip operator.

We prescribe marginal process tensors on the selected time-step sets
\(J_1:=\{0,1\}\), \(J_2:=\{1,2\}\), and \(J_3:=\{0,2\}\), and write
\(\Lambda:=\{J_1,J_2,J_3\}\).

The first marginal process tensor \(\mathcal T_{J_1}\) is a one-step process
with open Hilbert space
\begin{equation}
\mathcal H_{J_1}
=
\mathcal H_{i_1}\otimes \mathcal H_{o_0}\otimes \mathcal H_{i_0}.
\end{equation}
It prepares \(\rho_0\) at time \(0\) and transmits the output of the
time-\(0\) control to time \(1\). Thus, for every admissible control
operation \(\mathcal A_0\),
\begin{equation}
\mathcal T_{J_1}[\mathcal A_0]
=
\mathcal A_0(\rho_0).
\end{equation}

The second marginal process tensor \(\mathcal T_{J_2}\) is a one-step
process with open Hilbert space
\begin{equation}
\mathcal H_{J_2}
=
\mathcal H_{i_2}\otimes \mathcal H_{o_1}\otimes \mathcal H_{i_1}.
\end{equation}
It prepares \(\rho_0\) at time \(1\) and transmits the output of the
time-\(1\) control to time \(2\). Hence, for every admissible control
operation \(\mathcal A_1\),
\begin{equation}
\mathcal T_{J_2}[\mathcal A_1]
=
\mathcal A_1(\rho_0).
\end{equation}

The third marginal process tensor \(\mathcal T_{J_3}\) is a one-step process
with open Hilbert space
\begin{equation}
\mathcal H_{J_3}
=
\mathcal H_{i_2}\otimes \mathcal H_{o_0}\otimes \mathcal H_{i_0}.
\end{equation}
It prepares \(\rho_0\) at time \(0\), sends the output of the time-\(0\)
control to time \(2\), and applies the bit-flip channel. Therefore,
\begin{equation}
\mathcal T_{J_3}[\mathcal A_0]
=
\mathcal X\!\left(\mathcal A_0(\rho_0)\right)
=
X\mathcal A_0(\rho_0)X.
\end{equation}
Each of the three prescribed marginals is a valid one-step process tensor.

We now verify local compatibility. First, \(J_1\cap J_3=\{0\}\). Both
marginals reduce to the same initial state on the overlap:
\begin{equation}
\mathcal M_{\{0\}}(\mathcal T_{J_1})
=
\rho_0
=
\mathcal M_{\{0\}}(\mathcal T_{J_3}).
\end{equation}

Second, \(J_1\cap J_2=\{1\}\). Closing the time-\(0\) slot of
\(\mathcal T_{J_1}\) with the identity control \(\mathcal I_0\) gives
\begin{equation}
\mathcal M_{\{1\}}(\mathcal T_{J_1})
=
\mathcal T_{J_1}[\mathcal I_0]
=
\rho_0.
\end{equation}
On the other hand, the one-time marginal of \(\mathcal T_{J_2}\) at time
\(1\) is its prepared input state,
\begin{equation}
\mathcal M_{\{1\}}(\mathcal T_{J_2})
=
\rho_0.
\end{equation}
Thus,
\begin{equation}
\mathcal M_{\{1\}}(\mathcal T_{J_1})
=
\mathcal M_{\{1\}}(\mathcal T_{J_2}).
\end{equation}

Third, \(J_2\cap J_3=\{2\}\). Closing the time-\(1\) slot of
\(\mathcal T_{J_2}\) gives
\begin{equation}
\mathcal M_{\{2\}}(\mathcal T_{J_2})
=
\mathcal T_{J_2}[\mathcal I_1]
=
\rho_0.
\end{equation}
Similarly, closing the time-\(0\) slot of \(\mathcal T_{J_3}\) gives
\begin{equation}
\mathcal M_{\{2\}}(\mathcal T_{J_3})
=
\mathcal T_{J_3}[\mathcal I_0]
=
\mathcal X(\rho_0)
=
X\rho_0X
=
\rho_0.
\end{equation}
Therefore,
\begin{equation}
\mathcal M_{\{2\}}(\mathcal T_{J_2})
=
\mathcal M_{\{2\}}(\mathcal T_{J_3}).
\end{equation}
Hence the family
\(\boldsymbol{\mathcal T}:=\{\mathcal T_{J_1},\mathcal T_{J_2},
\mathcal T_{J_3}\}\) is locally compatible.

Nevertheless, this locally compatible family is not globally compatible.
That is, there is no two-step process tensor \(\mathcal T_{2:0}\) such that
\begin{equation}
\mathcal M_{J_1}(\mathcal T_{2:0})
=
\mathcal T_{J_1},
\qquad
\mathcal M_{J_2}(\mathcal T_{2:0})
=
\mathcal T_{J_2},
\qquad
\mathcal M_{J_3}(\mathcal T_{2:0})
=
\mathcal T_{J_3}.
\end{equation}

Suppose, for contradiction, that such a global process tensor
$\mathcal T_{2:0}$ exists, and let
\begin{equation}
\Upsilon_{2:0}:=J(\mathcal T_{2:0})\succeq0
\end{equation}
be its Choi operator. From the deterministic-comb causality condition
and the prescribed marginal on $J_1=\{0,1\}$, we have
\begin{equation}
\operatorname{Tr}_{i_2}\Upsilon_{2:0}
=
\mathbb I_{o_1}\otimes
\Phi^+_{i_1o_0}\otimes
(\rho_0)_{i_0},
\end{equation}
where $\Phi^+$ is the unnormalized Choi operator of the identity
channel.

The operator $\Phi^+_{i_1o_0}$ has rank one. For a positive operator,
if one of its marginals is supported on a subspace $S$, then the
operator itself is supported on $S$ in the corresponding subsystem.
Indeed, its expectation on $S^\perp$ vanishes, and positivity also
forces the associated off-diagonal blocks to vanish. Consequently,
there exists an operator $Q_{i_2o_1i_0}\succeq0$ such that
\begin{equation}
\Upsilon_{2:0}
=
\Phi^+_{i_1o_0}\otimes Q_{i_2o_1i_0},
\end{equation}
where the normalization factor is absorbed into $Q$.

The marginal on $J_2=\{1,2\}$ is obtained by closing the time-$0$
slot with the identity control. Hence
\begin{align}
\mathcal M_{J_2}(\Upsilon_{2:0})
&=
\operatorname{Tr}_{i_0o_0}
\left[
\left(
\Phi^+_{o_0i_0}\otimes \mathbb I_{i_2o_1i_1}
\right)
\Upsilon_{2:0}
\right]
\nonumber\\
&=
Q_{i_2o_1i_1},
\end{align}
where $Q_{i_2o_1i_1}$ denotes $Q_{i_2o_1i_0}$ with the subsystem
$i_0$ relabeled as $i_1$. Under the
unnormalized Choi convention used here, no additional transpose or
dimension factor appears.

Since the prescribed $J_2$ marginal has Choi operator
\begin{equation}
\mathcal M_{J_2}(\Upsilon_{2:0})
=
\Phi^+_{i_2o_1}\otimes(\rho_0)_{i_1},
\end{equation}
it follows that
\begin{equation}
Q_{i_2o_1i_0}
=
\Phi^+_{i_2o_1}\otimes(\rho_0)_{i_0}.
\end{equation}
Therefore, any global extension satisfying the $J_1$ and $J_2$
marginal conditions must have Choi operator
\begin{equation}
\Upsilon_{2:0}
=
\Phi^+_{i_2o_1}
\otimes
\Phi^+_{i_1o_0}
\otimes
(\rho_0)_{i_0}.
\end{equation}

Closing the time-$1$ slot with the identity control now gives
\begin{align}
\mathcal M_{J_3}(\Upsilon_{2:0})
&=
\operatorname{Tr}_{i_1o_1}
\left[
\left(
\Phi^+_{o_1i_1}\otimes \mathbb I_{i_2o_0i_0}
\right)
\Upsilon_{2:0}
\right]
\nonumber\\
&=
\Phi^+_{i_2o_0}\otimes(\rho_0)_{i_0}.
\end{align}
Thus, the induced channel from $o_0$ to $i_2$ is the identity channel,
and therefore
\begin{equation}
\mathcal M_{J_3}(\mathcal T_{2:0})[\mathcal A_0]
=
\mathcal A_0(\rho_0)
\end{equation}
for every admissible time-$0$ control operation $\mathcal A_0$.

This contradicts the prescribed marginal $\mathcal T_{J_3}$, which
requires
\begin{equation}
\mathcal T_{J_3}[\mathcal A_0]
=
\mathcal X\!\left(\mathcal A_0(\rho_0)\right).
\end{equation}
Indeed, choose $\mathcal A_0$ to be the replacement operation
preparing $\lvert0\rangle\langle0\rvert$. The global extension would
then imply
\begin{equation}
\mathcal M_{J_3}(\mathcal T_{2:0})[\mathcal A_0]
=
\lvert0\rangle\langle0\rvert,
\end{equation}
whereas the prescribed marginal gives
\begin{equation}
\mathcal T_{J_3}[\mathcal A_0]
=
X\lvert0\rangle\langle0\rvert X
=
\lvert1\rangle\langle1\rvert.
\end{equation}
This is a contradiction.

This contradiction is not detected by the one-time overlaps. Indeed,
the overlap tests only see the maximally mixed state, and
\begin{equation}
\mathcal X(\rho_0)
=
X\rho_0X
=
\rho_0.
\end{equation}
Thus, all one-time marginals agree, even though the three process
tensors do not admit a common global extension.

\section{Irreducibility to the channel marginal problem under channelization}
\label{app:ptmp-not-cmp}

\begin{proof}
Consider a classical bit process embedded in qubit systems on the time set
\(K=\{0,1,2,3\}\). All classical variables are encoded in the computational
basis of the corresponding qubit Hilbert spaces.

Let \(J_1:=\{0,2,3\}\) and \(J_2:=\{1,2,3\}\). We prescribe two
process tensor marginals \(\mathcal T_{J_1}\) and \(\mathcal T_{J_2}\).

Write \(P_r:=|r\rangle\langle r|\), with superscripts identifying the
tensor legs.

The first marginal \(\mathcal T_{J_1}\) samples a uniformly random bit
\(R\in\{0,1\}\), supplies \(|R\rangle\langle R|\) on \(i_0\),
supplies \(|0\rangle\langle0|\) on \(i_2\), and outputs
\(|R\rangle\langle R|\) on \(i_3\). The corresponding joint classical
state is
\begin{equation}
\frac{1}{2}\sum_{R=0}^1
|R\rangle\langle R|_{i_3}
\otimes
|0\rangle\langle 0|_{i_2}
\otimes
|R\rangle\langle R|_{i_0}.
\end{equation}
It discards the returned control-output systems \(o_0\) and \(o_2\).
Its complete process-tensor Choi operator is
\begin{equation}
\Upsilon_{J_1}
=\frac12\sum_{r=0}^{1}
P_r^{i_3}\otimes I_{o_2}\otimes P_0^{i_2}
\otimes I_{o_0}\otimes P_r^{i_0}.
\end{equation}

The second marginal \(\mathcal T_{J_2}\) samples a uniformly random bit
\(S\in\{0,1\}\), supplies \(|S\rangle\langle S|\) on \(i_1\), and
supplies \(|0\rangle\langle0|\) on \(i_2\). It then dephases the returned
system \(o_2\) in the computational basis. If the resulting value is
\(x\in\{0,1\}\), the process outputs
\(|S\oplus x\rangle\langle S\oplus x|\) on \(i_3\). The corresponding
joint classical state is
\begin{equation}
\frac{1}{2}\sum_{S=0}^1
|S\oplus x\rangle\langle S\oplus x|_{i_3}
\otimes
|0\rangle\langle 0|_{i_2}
\otimes
|S\rangle\langle S|_{i_1}.
\end{equation}
It ignores the returned control-output system \(o_1\). Its complete
process-tensor Choi operator is
\begin{equation}
\Upsilon_{J_2}
=\frac12\sum_{s,x=0}^{1}
P_{s\oplus x}^{i_3}\otimes P_x^{o_2}
\otimes P_0^{i_2}\otimes I_{o_1}\otimes P_s^{i_1}.
\end{equation}
The preceding classical state expressions describe the prepared
computational-basis states; the complete process-tensor Choi operators also
include the control-output legs. In the second classical expression, \(x\)
labels a fixed computational-basis input on \(o_2\), whereas the complete
Choi operator sums over both \(s\) and \(x\). The identity factors on
discarded control-output legs implement the trace operation in the
unnormalized Choi convention.

Both \(\mathcal T_{J_1}\) and \(\mathcal T_{J_2}\) are valid process
tensors. They are classical stochastic combs embedded into qubit systems:
the relevant control-output legs are dephased in the computational basis,
the stated classical stochastic rule is applied, and the resulting map is
extended linearly to arbitrary CP controls. Equivalently, their Choi
operators are diagonal positive operators satisfying the deterministic-comb
normalization constraints of Eq.~\eqref{eq:pt-causal-trace-constraints}.

We first verify local compatibility. The overlap is
\(J_1\cap J_2=\{2,3\}\). Let
\(\Phi_j^+:=J(\mathcal I_j)\) denote the unnormalized Choi operator of
the identity channel at time \(j\). The two identity-channel contractions
are
\begin{align}
\mathcal M_{\{2,3\}}(\Upsilon_{J_1})
&=\operatorname{Tr}_{i_0o_0}
\!\left[\Phi_0^+\Upsilon_{J_1}\right]
=\frac{I_{i_3}}{2}\otimes I_{o_2}\otimes P_0^{i_2},
\\
\mathcal M_{\{2,3\}}(\Upsilon_{J_2})
&=\operatorname{Tr}_{i_1o_1}
\!\left[\Phi_1^+\Upsilon_{J_2}\right]
=\frac{I_{i_3}}{2}\otimes I_{o_2}\otimes P_0^{i_2}.
\end{align}
Identities on untouched tensor factors are understood. In the second
contraction, the equality uses \(\sum_s P_{s\oplus x}=I\) for each \(x\)
and \(\sum_xP_x=I\).
Thus
\begin{equation}
\mathcal M_{\{2,3\}}(\mathcal T_{J_1})
=
\mathcal M_{\{2,3\}}(\mathcal T_{J_2}),
\end{equation}
so the prescribed marginals are locally compatible.

We now show that they are not globally compatible as causally ordered
process tensors. Assume, for contradiction, that there exists a causally
ordered global process tensor \(\mathcal T_{3:0}\) such that
\begin{equation}
\mathcal M_{J_1}(\mathcal T_{3:0})=\mathcal T_{J_1},
\qquad
\mathcal M_{J_2}(\mathcal T_{3:0})=\mathcal T_{J_2}.
\end{equation}
We prove incompatibility directly at the Choi-operator level, thereby keeping
the identity links at the omitted slots. Let
\(\Upsilon:=J(\mathcal T_{3:0})\), with tensor-factor order
\(i_3o_2i_2o_1i_1o_0i_0\). For
\(a,b,c,x\in\{0,1\}\), introduce the shorthand
\begin{equation}
 |c,x;b,a\rangle
 :=
 |c\rangle_{i_3}|x\rangle_{o_2}|0\rangle_{i_2}
 |b\rangle_{o_1}|b\rangle_{i_1}
 |a\rangle_{o_0}|a\rangle_{i_0},
\end{equation}
and let \(\Upsilon_{cx}\) be the corresponding principal submatrix,
\begin{equation}
 [\Upsilon_{cx}]_{ba,b'a'}
 :=
 \langle c,x;b,a|\Upsilon|c,x;b',a'\rangle .
\end{equation}
Each \(\Upsilon_{cx}\) is positive semidefinite because
\(\Upsilon\succeq0\).

On the four-dimensional space with ordered basis
\(\{|00\rangle,|01\rangle,|10\rangle,|11\rangle\}\), where the first bit
is \(b\) and the second is \(a\), define
\begin{equation}
 |u_a\rangle:=\sum_{b=0}^1|ba\rangle,
 \qquad
 |v_b\rangle:=\sum_{a=0}^1|ba\rangle .
\end{equation}
In the \(o_ji_j\) tensor order, these identity-channel Choi operators have
the form
\begin{equation}
 \Phi_j^+
 =
 |\Phi_j^+\rangle\langle\Phi_j^+|,
 \qquad
 |\Phi_j^+\rangle
 :=
 \sum_{r=0}^1 |r\rangle_{o_j}|r\rangle_{i_j},
\end{equation}
Contracting the omitted time-\(1\) slot of the first marginal with
\(\Phi_1^+\), and the omitted time-\(0\) slot of the second marginal with
\(\Phi_0^+\), gives
\begin{align}
 \langle u_a|\Upsilon_{cx}|u_a\rangle
 &=
 \sum_{b,b'=0}^1
 \langle c,x;b,a|\Upsilon|c,x;b',a\rangle
 =\frac12\,\delta_{c,a},
 \label{eq:appG-link-J1}\\
 \langle v_b|\Upsilon_{cx}|v_b\rangle
 &=
 \sum_{a,a'=0}^1
 \langle c,x;b,a|\Upsilon|c,x;b,a'\rangle
 =\frac12\,\delta_{c,b\oplus x}.
 \label{eq:appG-link-J2}
\end{align}
The sums over \(b,b'\) and \(a,a'\) are precisely the contractions
with the identity-channel Choi operators
\begin{equation}
 \Phi_1^+
 =
 \sum_{b,b'=0}^1 |bb\rangle\langle b'b'|,
 \qquad
 \Phi_0^+
 =
 \sum_{a,a'=0}^1 |aa\rangle\langle a'a'|,
\end{equation}
respectively.

The last deterministic-comb causality condition is
\begin{equation}
 \operatorname{Tr}_{i_3}\Upsilon
 =
 \mathbb I_{o_2}\otimes\Upsilon_{2:0}.
\end{equation}
Restricting this equality to the four-dimensional subspace defined above,
tracing over \(i_3\) gives the sum over \(c\), while the factor
\(\mathbb I_{o_2}\) makes the resulting diagonal block independent of
\(x\). Hence
\begin{equation}
 \sum_{c=0}^1\Upsilon_{c0}
 =
 \sum_{c=0}^1\Upsilon_{c1}.
 \label{eq:appG-causal-matrix}
\end{equation}
Consequently, for every Hermitian four-by-four matrix \(H\),
\begin{equation}
 \sum_{c=0}^1
 \operatorname{Tr}\!\left[
 H(\Upsilon_{c0}-\Upsilon_{c1})
 \right]
 =0.
 \label{eq:appG-causal-block}
\end{equation}

Set
\begin{equation}
 \alpha:=\frac{1-\sqrt2}{8},
 \qquad
 \beta:=\frac{1+\sqrt2}{8},
 \qquad
 \gamma_{pq}:=
 \begin{cases}
   \alpha,&p=q,\\
   \beta,&p\ne q,
 \end{cases}
\end{equation}
and choose
\begin{equation}
 H:=\frac18
 \begin{pmatrix}
  1& 0& 0&-1\\
  0&-1& 1& 0\\
  0& 1&-1& 0\\
 -1& 0& 0& 1
 \end{pmatrix}.
\end{equation}
For each \(c,x\), define
\begin{equation}
 W_{cx}
 :=
 \sum_{a=0}^1\gamma_{c,a}|u_a\rangle\langle u_a|
 +\sum_{b=0}^1\gamma_{c,b\oplus x}|v_b\rangle\langle v_b|
 +(-1)^xH.
 \label{eq:appG-Wcx}
\end{equation}
Although \(\alpha<0\), every \(W_{cx}\) is positive semidefinite. Indeed,
a direct four-by-four calculation gives
\begin{equation}
 W_{cx}=|w_{cx}\rangle\langle w_{cx}|,
\end{equation}
where
\begin{align}
 |w_{00}\rangle
 &=
 \frac{2-\sqrt2}{4}|00\rangle
 -\frac{\sqrt2}{4}(|01\rangle+|10\rangle)
 -\frac{2+\sqrt2}{4}|11\rangle,\\
 |w_{cx}\rangle
 &=
 \bigl(X^{c\oplus x}\otimes X^c\bigr)|w_{00}\rangle .
\end{align}
Here the first tensor factor acts on the \(b\) label and the second on the
\(a\) label.

Since \(W_{cx}=|w_{cx}\rangle\langle w_{cx}|\) and
\(\Upsilon_{cx}\succeq0\),
\begin{equation}
 \operatorname{Tr}(W_{cx}\Upsilon_{cx})
 =
 \langle w_{cx}|\Upsilon_{cx}|w_{cx}\rangle
 \ge 0.
\end{equation}
Hence
\begin{equation}
 0
 \le
 \sum_{c,x=0}^1\operatorname{Tr}(W_{cx}\Upsilon_{cx}).
 \label{eq:appG-positive-sum}
\end{equation}
On the other hand, substituting Eqs.~\eqref{eq:appG-link-J1} and
\eqref{eq:appG-link-J2} into Eq.~\eqref{eq:appG-Wcx}, and using
Eq.~\eqref{eq:appG-causal-block} for the terms containing \(H\), yields
\begin{align}
 \sum_{c,x=0}^1\operatorname{Tr}(W_{cx}\Upsilon_{cx})
 &=
 \frac12\sum_{c,x,a=0}^1
 \gamma_{c,a}\delta_{c,a}
 +
 \frac12\sum_{c,x,b=0}^1
 \gamma_{c,b\oplus x}\delta_{c,b\oplus x}
 \notag\\
 &\quad+
 \sum_{c=0}^1
 \operatorname{Tr}\!\left[H(\Upsilon_{c0}-\Upsilon_{c1})\right]
 \\
 &=2\alpha+2\alpha+0
 \\
 &=\frac{1-\sqrt2}{2}<0,
\end{align}
contradicting Eq.~\eqref{eq:appG-positive-sum}. Hence no causally ordered
global process tensor can have both prescribed temporal marginals.

It remains to show that the obstruction disappears after forgetting the
temporal ordering. We now regard each process tensor marginal as an ordinary
channel by grouping all selected \(o\)-legs as channel inputs and all
selected \(i\)-legs as channel outputs. Thus \(\mathcal T_{J_1}\) is viewed
as a channel from \(o_2o_0\) to \(i_3i_2i_0\), and \(\mathcal T_{J_2}\) is
viewed as a channel from \(o_2o_1\) to \(i_3i_2i_1\).

Define a global quantum channel
\(\Phi:o_2o_1o_0\to i_3i_2i_1i_0\) by the measure-and-prepare rule
\begin{equation}
\Phi(\rho_{o_2o_1o_0})
=
\sum_{x=0}^1
\operatorname{Tr}\!\left[
\left(
|x\rangle\langle x|_{o_2}\otimes \mathbb I_{o_1o_0}
\right)
\rho_{o_2o_1o_0}
\right]
\tau_x,
\end{equation}
where
\begin{equation}
\tau_x
:=
\frac{1}{2}\sum_{R=0}^1
|R\rangle\langle R|_{i_3}
\otimes
|0\rangle\langle 0|_{i_2}
\otimes
|R\oplus x\rangle\langle R\oplus x|_{i_1}
\otimes
|R\rangle\langle R|_{i_0}.
\end{equation}
This is a valid CPTP channel.

Since \(\Phi\) is independent of \(o_1\) and \(o_0\), it is
no-signaling from \(o_1\) to \(i_3 i_2 i_0\) and from \(o_0\)
to \(i_3 i_2 i_1\). Hence both reduced channels are well
defined in the sense of Ref.~\cite{Hsieh:2021ewp}.

Taking the marginal of \(\Phi\) on \(J_1\), i.e., keeping outputs
\(i_3,i_2,i_0\) and inputs \(o_2,o_0\), gives
\begin{equation}
\operatorname{Tr}_{i_1}\tau_x
=
\frac{1}{2}\sum_{R=0}^1
|R\rangle\langle R|_{i_3}
\otimes
|0\rangle\langle 0|_{i_2}
\otimes
|R\rangle\langle R|_{i_0}.
\end{equation}
This is independent of \(x\) and is exactly the channelization of
\(\mathcal T_{J_1}\).

Taking the marginal of \(\Phi\) on \(J_2\), i.e., keeping outputs
\(i_3,i_2,i_1\) and inputs \(o_2,o_1\), gives
\begin{equation}
\operatorname{Tr}_{i_0}\tau_x
=
\frac{1}{2}\sum_{R=0}^1
|R\rangle\langle R|_{i_3}
\otimes
|0\rangle\langle 0|_{i_2}
\otimes
|R\oplus x\rangle\langle R\oplus x|_{i_1}.
\end{equation}
Let \(S:=R\oplus x\). Since \(R\) is uniformly random, \(S\) is uniformly
random for every fixed \(x\), and \(R=S\oplus x\). Therefore
\begin{equation}
\operatorname{Tr}_{i_0}\tau_x
=
\frac{1}{2}\sum_{S=0}^1
|S\oplus x\rangle\langle S\oplus x|_{i_3}
\otimes
|0\rangle\langle 0|_{i_2}
\otimes
|S\rangle\langle S|_{i_1}.
\end{equation}
This is exactly the channelization of \(\mathcal T_{J_2}\).

Hence the channelized marginals are compatible as ordinary channels, even
though the original process tensor marginals are incompatible as causally
ordered process tensors. Therefore PTMP compatibility cannot be reduced to
CMP compatibility by channelization.
\end{proof}

\section{Proof of Theorem~\ref{thm:discrimination-advantage}}
\label{app:proof-operational-interpretation}

We first show that every completely positive multi-time control operation can be embedded into a deterministic multi-time process with a finite-dimensional auxiliary readout system.

\begin{lemma}[Deterministic embedding of a completely positive multi-time operation]
\label{lem:deterministic-completion}
Let \(\mathbf A\) be a completely positive multi-time control operation with fixed interactive input-output interfaces. Then there exist a number \(s>0\), a completely positive multi-time control operation \(\mathbf C\), and a two-dimensional auxiliary readout system \(R\) such that \(s\mathbf A+\mathbf C\) is deterministic. Moreover,
\begin{equation}
    \mathcal P
    :=
    s\mathbf A\otimes |1\rangle\langle1|_R
    +
    \mathbf C\otimes |0\rangle\langle0|_R
\end{equation}
is a deterministic multi-time process with auxiliary readout system \(R\).
\end{lemma}

\begin{proof}
Choose a deterministic quantum comb with the same interactive input-output interfaces as \(\mathbf A\) and with a strictly positive Choi operator \(\Omega\succ0\). Since \(J(\mathbf A)\succeq0\), one can choose \(s>0\) sufficiently small such that
\begin{equation}
    sJ(\mathbf A)\preceq\Omega.
\end{equation}
Define \(\mathbf C\) by
\begin{equation}
    J(\mathbf C)
    :=
    \Omega-sJ(\mathbf A)
    \succeq0.
\end{equation}
Thus \(\mathbf C\) is completely positive, and
\begin{equation}
    J(s\mathbf A+\mathbf C)
    =
    sJ(\mathbf A)+J(\mathbf C)
    =
    \Omega.
\end{equation}
Since \(\Omega\) is the Choi operator of a deterministic quantum comb, \(s\mathbf A+\mathbf C\) is deterministic.

Introduce a two-dimensional auxiliary readout system \(R\) and define
\begin{equation}
    \mathcal P
    :=
    s\mathbf A\otimes |1\rangle\langle1|_R
    +
    \mathbf C\otimes |0\rangle\langle0|_R.
\end{equation}
Its Choi operator is
\begin{equation}
    J(\mathcal P)
    =
    sJ(\mathbf A)\otimes |1\rangle\langle1|_R
    +
    J(\mathbf C)\otimes |0\rangle\langle0|_R
    \succeq0,
\end{equation}
and satisfies
\begin{equation}
    \operatorname{Tr}_R J(\mathcal P)
    =
    sJ(\mathbf A)+J(\mathbf C)
    =
    \Omega.
\end{equation}

Suppose that the control comb has \(n\geq1\) interactive interfaces, and write \(\Omega_{n-1:0}:=\Omega\). Since \(\Omega\) is deterministic, there exist lower-order combs \(\Omega_{j:0}\) satisfying
\begin{equation}
    \operatorname{Tr}_{o_j}\Omega_{j:0}
    =
    \mathbb I_{i_j}\otimes\Omega_{j-1:0},
    \qquad
    j=0,\ldots,n-1,
\end{equation}
where \(\Omega_{-1:0}:=1\). The auxiliary readout system \(R\) is appended to the final output \(o_{n-1}\) and does not introduce an additional interactive slot. Consequently,
\begin{equation}
\begin{split}
    \operatorname{Tr}_{o_{n-1}R}J(\mathcal P)
    &=
    \operatorname{Tr}_{o_{n-1}}
    \left[
        \operatorname{Tr}_R J(\mathcal P)
    \right]
    \\
    &=
    \operatorname{Tr}_{o_{n-1}}\Omega_{n-1:0}
    \\
    &=
    \mathbb I_{i_{n-1}}
    \otimes
    \Omega_{n-2:0}.
\end{split}
\end{equation}
The remaining recursive normalization constraints are inherited unchanged from \(\Omega\). Together with \(J(\mathcal P)\succeq0\), this shows that \(J(\mathcal P)\) satisfies the characterization of a deterministic quantum comb recalled in Sec.~\ref{sec:iiib}. Therefore, \(\mathcal P\) is deterministic.
\end{proof}

We now prove Theorem~\ref{thm:discrimination-advantage}.

\begin{proof}
Suppose first that there exists a multi-time process-discrimination task \(D\) such that
\begin{equation}
    P(D,\boldsymbol{\mathcal T})
    >
    P_{\mathfrak C}(D).
\end{equation}
If \(\boldsymbol{\mathcal T}\in\mathfrak C\), then the definition of the compatible benchmark gives
\begin{equation}
    P(D,\boldsymbol{\mathcal T})
    \leq
    P_{\mathfrak C}(D),
\end{equation}
which is a contradiction. Hence, \(\boldsymbol{\mathcal T}\) is incompatible.

Conversely, suppose that \(\boldsymbol{\mathcal T}\notin\mathfrak C\). By Theorem~\ref{thm:process-tensor-incompatibility-witness}, for every \(\alpha\in\Lambda\) there exist finitely many Hermitian final-output operators \(F_\alpha^{(i)}\) and completely positive multi-time control operations \(\mathbf A_\alpha^{(i)}\), indexed by \(i=1,\ldots,n_\alpha\), such that, upon defining
\begin{equation}
    L(\boldsymbol{\mathcal X})
    :=
    \sum_{\alpha\in\Lambda}
    \sum_{i=1}^{n_\alpha}
    \operatorname{Tr}\left[
        F_\alpha^{(i)}
        \mathcal X_\alpha
        \left[
            \mathbf A_\alpha^{(i)}
        \right]
    \right],
\end{equation}
one has
\begin{equation}
    L(\boldsymbol{\mathcal T})
    >
    \max_{\boldsymbol{\Gamma}\in\mathfrak C}
    L(\boldsymbol{\Gamma}).
\end{equation}
Set
\begin{equation}
    \delta
    :=
    L(\boldsymbol{\mathcal T})
    -
    \max_{\boldsymbol{\Gamma}\in\mathfrak C}
    L(\boldsymbol{\Gamma})
    >0.
\end{equation}

For each \(\alpha\in\Lambda\), fix a two-dimensional auxiliary readout system \(R_\alpha\). Applying Lemma~\ref{lem:deterministic-completion} to each \(\mathbf A_\alpha^{(i)}\), using \(R_\alpha\) as the auxiliary readout system, gives numbers \(s_{\alpha i}>0\), completely positive multi-time control operations \(\mathbf C_\alpha^{(i)}\), and deterministic processes
\begin{equation}
    \mathcal P_\alpha^{(i)}
    :=
    s_{\alpha i}\mathbf A_\alpha^{(i)}
        \otimes |1\rangle\langle1|_{R_\alpha}
    +
    \mathbf C_\alpha^{(i)}
        \otimes |0\rangle\langle0|_{R_\alpha}.
\end{equation}
For fixed \(\alpha\), these processes have the same interactive input-output interfaces and the same auxiliary readout system \(R_\alpha\).

For any family of valid marginal process tensors \(\boldsymbol{\mathcal X}\), define the final output state
\begin{equation}
    \rho_{\alpha i}(\boldsymbol{\mathcal X})
    :=
    \left(
        \mathcal X_\alpha
        \otimes
        \mathcal I_{R_\alpha}
    \right)
    \left[
        \mathcal P_\alpha^{(i)}
    \right].
\end{equation}
Since \(\mathcal P_\alpha^{(i)}\) is deterministic,
\begin{equation}
    \operatorname{Tr}
    \rho_{\alpha i}(\boldsymbol{\mathcal X})
    =
    1.
\end{equation}
Moreover,
\begin{equation}
\begin{split}
    \rho_{\alpha i}(\boldsymbol{\mathcal X})
    &=
    s_{\alpha i}
    \mathcal X_\alpha
    \left[
        \mathbf A_\alpha^{(i)}
    \right]
    \otimes |1\rangle\langle1|_{R_\alpha}
    \\
    &\quad+
    \mathcal X_\alpha
    \left[
        \mathbf C_\alpha^{(i)}
    \right]
    \otimes |0\rangle\langle0|_{R_\alpha}.
\end{split}
\end{equation}
Define
\begin{equation}
    \widehat F_\alpha^{(i)}
    :=
    \frac{1}{s_{\alpha i}}
    F_\alpha^{(i)}
    \otimes |1\rangle\langle1|_{R_\alpha}.
\end{equation}
The orthogonality of \(|0\rangle_{R_\alpha}\) and \(|1\rangle_{R_\alpha}\) gives
\begin{equation}
    \operatorname{Tr}\left[
        \widehat F_\alpha^{(i)}
        \rho_{\alpha i}(\boldsymbol{\mathcal X})
    \right]
    =
    \operatorname{Tr}\left[
        F_\alpha^{(i)}
        \mathcal X_\alpha
        \left[
            \mathbf A_\alpha^{(i)}
        \right]
    \right],
\end{equation}
and hence
\begin{equation}
    L(\boldsymbol{\mathcal X})
    =
    \sum_{\alpha\in\Lambda}
    \sum_{i=1}^{n_\alpha}
    \operatorname{Tr}\left[
        \widehat F_\alpha^{(i)}
        \rho_{\alpha i}(\boldsymbol{\mathcal X})
    \right].
\end{equation}

Choose strictly positive probability distributions \(\{p_\alpha\}_{\alpha\in\Lambda}\) and \(\{\overline q_\alpha^{(i)}\}_{i=1}^{n_\alpha}\), so that
\begin{equation}
    \sum_{\alpha\in\Lambda}p_\alpha=1,
    \qquad
    \sum_{i=1}^{n_\alpha}
    \overline q_\alpha^{(i)}=1.
\end{equation}
For every \(\alpha\) and \(i\), choose \(\nu_\alpha^{(i)}\) sufficiently large that
\begin{equation}
    H_\alpha^{(i)}
    :=
    \frac{
        \widehat F_\alpha^{(i)}
    }{
        p_\alpha\overline q_\alpha^{(i)}
    }
    +
    \nu_\alpha^{(i)}
    \mathbb I_{Y_\alpha\otimes R_\alpha}
    \succ0.
\end{equation}
Using the normalization of \(\rho_{\alpha i}(\boldsymbol{\mathcal X})\), we obtain
\begin{equation}
\begin{split}
    &\sum_{\alpha\in\Lambda}
    p_\alpha
    \sum_{i=1}^{n_\alpha}
    \overline q_\alpha^{(i)}
    \operatorname{Tr}\left[
        H_\alpha^{(i)}
        \rho_{\alpha i}(\boldsymbol{\mathcal X})
    \right]
    \\
    &\qquad=
    L(\boldsymbol{\mathcal X})+c,
\end{split}
\end{equation}
where
\begin{equation}
    c
    :=
    \sum_{\alpha\in\Lambda}
    p_\alpha
    \sum_{i=1}^{n_\alpha}
    \overline q_\alpha^{(i)}
    \nu_\alpha^{(i)}
\end{equation}
is independent of \(\boldsymbol{\mathcal X}\).

Since \(\Lambda\) and all index sets are finite, and all Hilbert spaces are finite-dimensional, one can choose \(\kappa>0\) sufficiently small such that
\begin{equation}
    \kappa
    \sum_{i=1}^{n_\alpha}
    H_\alpha^{(i)}
    \prec
    \mathbb I_{Y_\alpha\otimes R_\alpha}
\end{equation}
for every \(\alpha\). Define
\begin{equation}
    M_\alpha^{(i)}
    :=
    \kappa H_\alpha^{(i)},
    \qquad
    i=1,\ldots,n_\alpha,
\end{equation}
and
\begin{equation}
    M_\alpha^{(0)}
    :=
    \mathbb I_{Y_\alpha\otimes R_\alpha}
    -
    \sum_{i=1}^{n_\alpha}
    M_\alpha^{(i)}.
\end{equation}
Then
\begin{equation}
    \left\{
        M_\alpha^{(i)}
    \right\}_{i=0}^{n_\alpha}
\end{equation}
is a POVM, with every effect strictly positive.

For each \(\alpha\), choose a deterministic multi-time process \(\mathcal P_\alpha^{(0)}\) with the same interactive input-output interfaces and auxiliary readout system \(R_\alpha\), and set
\begin{equation}
    \rho_{\alpha0}(\boldsymbol{\mathcal X})
    :=
    \left(
        \mathcal X_\alpha
        \otimes
        \mathcal I_{R_\alpha}
    \right)
    \left[
        \mathcal P_\alpha^{(0)}
    \right].
\end{equation}

Let \(0<\epsilon<1\) and define
\begin{equation}
    q_\alpha^{(0)}
    :=
    \epsilon,
    \qquad
    q_\alpha^{(i)}
    :=
    (1-\epsilon)
    \overline q_\alpha^{(i)},
    \quad
    i=1,\ldots,n_\alpha.
\end{equation}
Together with the probabilities \(p_\alpha\), the processes \(\{\mathcal P_\alpha^{(i)}\}_{i=0}^{n_\alpha}\), and the POVM \(\{M_\alpha^{(i)}\}_{i=0}^{n_\alpha}\), these data define a valid multi-time process-discrimination task \(D\).

For any family of valid marginal process tensors \(\boldsymbol{\mathcal X}\), its success probability is
\begin{equation}
\begin{split}
    P(D,\boldsymbol{\mathcal X})
    &=
    (1-\epsilon)
    \sum_{\alpha\in\Lambda}
    p_\alpha
    \sum_{i=1}^{n_\alpha}
    \overline q_\alpha^{(i)}
    \operatorname{Tr}\left[
        M_\alpha^{(i)}
        \rho_{\alpha i}(\boldsymbol{\mathcal X})
    \right]
    \\
    &\quad+
    \epsilon
    \sum_{\alpha\in\Lambda}
    p_\alpha
    \operatorname{Tr}\left[
        M_\alpha^{(0)}
        \rho_{\alpha0}(\boldsymbol{\mathcal X})
    \right]
    \\
    &=
    (1-\epsilon)\kappa
    \left[
        L(\boldsymbol{\mathcal X})+c
    \right]
    +
    \epsilon B(\boldsymbol{\mathcal X}),
\end{split}
\end{equation}
where
\begin{equation}
    B(\boldsymbol{\mathcal X})
    :=
    \sum_{\alpha\in\Lambda}
    p_\alpha
    \operatorname{Tr}\left[
        M_\alpha^{(0)}
        \rho_{\alpha0}(\boldsymbol{\mathcal X})
    \right].
\end{equation}
Since \(M_\alpha^{(0)}\) is a POVM effect and \(\rho_{\alpha0}(\boldsymbol{\mathcal X})\) is normalized,
\begin{equation}
    0
    \leq
    B(\boldsymbol{\mathcal X})
    \leq
    1.
\end{equation}

For every \(\boldsymbol{\Gamma}\in\mathfrak C\),
\begin{equation}
\begin{split}
    P(D,\boldsymbol{\mathcal T})
    -
    P(D,\boldsymbol{\Gamma})
    &=
    (1-\epsilon)\kappa
    \left[
        L(\boldsymbol{\mathcal T})
        -
        L(\boldsymbol{\Gamma})
    \right]
    \\
    &\quad+
    \epsilon
    \left[
        B(\boldsymbol{\mathcal T})
        -
        B(\boldsymbol{\Gamma})
    \right]
    \\
    &\geq
    (1-\epsilon)\kappa\delta-\epsilon.
\end{split}
\end{equation}
Choose
\begin{equation}
    0<\epsilon
    <
    \frac{\kappa\delta}{1+\kappa\delta}.
\end{equation}
Then
\begin{equation}
    P(D,\boldsymbol{\mathcal T})
    >
    P(D,\boldsymbol{\Gamma})
\end{equation}
for every \(\boldsymbol{\Gamma}\in\mathfrak C\). Therefore,
\begin{equation}
    P(D,\boldsymbol{\mathcal T})
    >
    \max_{\boldsymbol{\Gamma}\in\mathfrak C}
    P(D,\boldsymbol{\Gamma})
    =
    P_{\mathfrak C}(D).
\end{equation}
This completes the proof.
\end{proof}

\end{document}